\documentclass[a4paper,11pt]{article}
\pdfoutput=1 % if your are submitting a pdflatex (i.e. if you have
\usepackage{jcappub} % for details on the use of the package, please
\usepackage[T1]{fontenc} % if needed
\usepackage{amssymb} 
\usepackage{subcaption}
\usepackage{multirow}
\usepackage{float}
\usepackage{forest}

\def\nn{\nonumber}

\def\R2{{\mathcal R}^2}

\def\btc{\begin{tcolorbox}}
\def\etc{\end{tcolorbox}}
\def\be{\begin{equation}}
\def\ee{\end{equation}}
\def\barr{\begin{array}{lr}}
\def\earr{\end{array}}
\def\bea{\begin{eqnarray}}
\def\eea{\end{eqnarray}}
\def\la{\langle}
\def\ra{\rangle}

\def\nn{\nonumber}

\def\s{\sigma}

\usepackage{xcolor}
\usepackage[dvipsnames]{xcolor}

\arxivnumber{}
\title{\boldmath Minkowski Functionals of the 21~cm Signal as a Probe of Primordial Features}

\author[a,b,1]{Kanan Virkar,}
\author[c,a,2]{Suvedha Suresh Naik,}
\author[a,b,3]{Pravabati Chingangbam}

\affiliation[a]{Indian Institute of Astrophysics, Koramangala II Block, Bangalore 560 034, India.}
\affiliation[b]{Department of Physics, Pondicherry University, R.V. Nagar, Kalapet, 605 014, Puducherry, India}
\affiliation[c]{Korea Institute for Advanced Study (KIAS), 85 Hoegiro, Dongdaemun-gu, Seoul, Republic of Korea-02455}

\emailAdd{kanan.virkar@iiap.res.in}
\emailAdd{suvedha@kias.re.kr}
\emailAdd{prava@iiap.res.in}

\abstract{
The redshifted 21~cm signal from the cosmic dawn and Epoch of Reionization (EoR) encodes important information about both astrophysical processes and primordial physics, such as inflation. In this work, we use morphological statistics to explore the sensitivity of the 21~cm signal to inflationary features and EoR dynamics simultaneously. Focusing on primordial features from particle production during inflation we generate semi-numerical simulations of the 21~cm signal across redshifts $5 \lesssim z \lesssim 35$, incorporating these features. Using Minkowski Functionals (MFs), we analyze the morphology of 21~cm fields: density, neutral hydrogen fraction, spin temperature, and brightness temperature. 
We demonstrate that MFs are highly sensitive to both the amplitude and scale of primordial features, capturing rich morphological information.
In particular, we show that MFs can robustly identify inflationary features and distinguish them from the standard model. We further  explore various EoR scenarios, and demonstrate that combining MFs across redshifts can disentangle the signatures of primordial features from EoR effects. This approach opens new avenues for probing inflation with upcoming 21~cm surveys.
}
\begin{document}
%version: {\today}
\maketitle
\flushbottom
%%%%%%%%%%%%%%%%%%%%%%%%%%%%%%%%%%%%%%%%%%%%%%%%%%%%%%%%%%%%%%%%%%%%%%
\section{Introduction}

%\red{
%Point 1: Overview and importance of the 21~cm signal, usefulness in studying the cosmic dawn, EoR, cite important literatures.\\
%Point 2: Usefulness of 21~cm signal for the study of primordial physics: inflation, dark matter, etc. \\
%Point 3: Quantifying the 21~cm signal, upcoming observations and their targets. Motivation to use morphology.\\
%Point 4: Minkowski functionals, application of MFs in the context of EoR, cite all relevant literatures.\\
%Point 5: Objective of the paper, important findings
%}

%\red{Introduction is not yet in good shape. The objectives of why we are carrying out this study is not spelled out clearly. You need to state what has been done previously by others, cite all relevant references, and then state what is new and important in this work. }

% Point 1: Overview and importance of the 21 cm signal
The first billion years after the Big Bang
is an important period in cosmology during which  the first stars, % and structures like 
galaxies and the large scale structure formed through hierarchical structure formation. These early luminous sources ionized the neutral hydrogen surrounding them, and by around $z \sim 6$ %the end of this period (around $z \sim 6$) 
almost all of the neutral hydrogen in the  universe was ionized. This period,  known as the epoch of reionization (EoR), spans from the cosmic dawn (when the first stars formed) to the end of reionization. Despite its cosmological significance, this epoch remains relatively unexplored observationally. However, recent observational efforts by modern telescopes like JWST  shed light on early galaxies up to redshift $z \sim 14$ \cite{2024Natur.633..318C}.  The high-redshift frontier has also been extended through observations of Lyman-$\alpha$ forest from high-redshift 
quasars that reveal properties of intervening neutral hydrogen clouds \cite{Fan_2006}. One of the most promising probes of this epoch is the 21~cm signal from neutral hydrogen. This signal is produced by the hyperfine transition in the ground state of neutral hydrogen in the radio band ($\lambda \sim 21$\,cm or $\nu \sim 1420$\,MHz).  Since neutral hydrogen is the most abundant baryonic component in the early universe, the 21~cm signal provides direct information about the thermal and ionisation state of the intergalactic medium (IGM), the properties of early luminous sources, and the underlying cosmology (see, e.g., \cite{Furlanetto:2006jb,Pritchard:2011xb} for reviews).

% Point 2: Usefulness of 21 cm signal for primordial physics
%Beyond probing astrophysical processes, the 21~cm signal offers a unique window into primordial physics. 
The distribution of neutral hydrogen during the cosmic dawn and EoR is sensitive to the seed initial density fluctuations set by inflation. 
A large class of inflationary models predict deviations from a nearly scale-invariant form of the 
primordial power spectrum, commonly referred to as `primordial features'~(see \cite{Chluba:2015bqa} for a review). The 21~cm signal then offers a unique window into primordial density fluctuations, besides its sensitivity to astrophysical processes, providing complementary probes of inflationary signatures alongside the CMB and large-scale structure observations.  
%The 21~cm signal thus provides a complementary probe of inflationary signatures alongside the CMB and large-scale structure observations. 
Additionally, the 21~cm signal has been proposed as a probe of dark matter properties (see, e.g.,   \cite{Zhao:2025ddy,Park:2025phj} and refs. therein), primordial non-Gaussianity \cite{Cooray:2008eb,Pillepich:2006fj,Munoz:2015eqa,Meerburg:2016zdz,Floss:2022grj}, and other beyond-standard-model physics.

In this work, we focus on bump-like features in the primordial power 
spectrum arising from particle production during 
inflation~\cite{Chung:1999ve,Barnaby:2009mc,Barnaby:2009dd,Pearce:2017bdc,Furuuchi:2015foh,Furuuchi:2020klq,Furuuchi:2020ery}. 
Such features can leave distinct imprints on the
observables such as CMB anisotropy \cite{Naik:2022mxn} and 
21~cm signal, as 
they modify the reionisation history and the distribution of 
structures depending on the scale at which the feature 
occurs~\cite{Naik:2022wej,Naik:2025mba}. However, in~\cite{Naik:2025mba} 
it was found that at specific scales -- namely the turnover scale -- bump 
models are indistinguishable from the fiducial model in their 
globally averaged 21~cm profile. It is therefore of importance to 
investigate whether other statistical measures can detect these 
features. As these features are theoretically well motivated, 
identifying their observational signatures is essential for probing the physics of inflation 
and the origin of primordial density fluctuations.

% Point 3: Quantifying the 21 cm signal, upcoming observations, motivation for morphology
Several ongoing and upcoming radio interferometers aim to 
detect the 21~cm signal from the EoR, including LOFAR~\cite{LOFAR2:2021A&A...652A..37E}, 
MWA~\cite{MWA2:2018PASA...35...33W}, HERA~\cite{DeBoer:2016tnn}, and the 
SKA~\cite{ska}. The SKA, in particular, will carry out 
the deepest observations of neutral hydrogen, tracing the evolution 
of cosmic structure over the redshift range $6 < z < 30$.  These observations will yield 
three-dimensional maps of the 21~cm brightness temperature, 
enabling detailed studies of the EoR morphology. Extracting cosmological and astrophysical information from 
these maps requires appropriate statistical tools.  The 21~cm signal from the EoR is expected to be highly 
non-Gaussian due to the patchy nature of reionisation; consequently, the globally averaged signal and power spectrum do not capture the full information content. Higher-order statistics, such as the bispectrum \cite{Pillepich:2006fj,Bharadwaj:2004sx,Shimabukuro:2015iqa,Majumdar:2017tdm,Watkinson:2018efd,Hutter:2019yta,Noble:2024uzl} and morphological descriptors, are therefore essential for fully  characterising the signal.

% Point 4: Minkowski functionals and their application to EoR
Minkowski functionals (MFs) \cite{Tomita:1986,Mecke:1994,Schmalzing:1997,Matsubara:2003} are a powerful set of morphological 
descriptors that characterise the geometry and topology of random 
fields. % \cite{adler}. 
In three dimensions, 
the four MFs quantify the volume fraction, surface area, mean 
curvature, and Euler characteristic of excursion sets at different 
thresholds. MFs have been extensively applied to study reionisation 
models using the 21~cm signal~\cite{Gleser_2006,yoshiura2016,Diao_2024,Bag:2018,Bag:2019,Ghara:2023efi}, 
extending beyond the genus~\cite{Lee_2008}.  Minkowski tensors~\cite{Ganesan:2017,Chingangbam:2017,Appleby:2018a,Appleby:2018b,Kapahtia:2017qrg,Kapahtia_2019,Kapahtia:2021eok} and Betti numbers~\cite{Park:2013,Kapahtia:2017qrg,Kapahtia_2019,Giri_2021,Kapahtia:2021eok,2019MNRAS.486.1523E,Elbers_2023} further provide additional information. These studies have demonstrated 
that MFs are sensitive to the ionisation topology and can distinguish 
between different reionisation scenarios. 
However, the imprints of primordial features on the morphology 
of the 21~cm signal have not yet been explored. Since bump-like 
features in the primordial power spectrum enhance density fluctuations 
at specific scales, they are expected to modify the clustering of 
matter and consequently the morphology of the 21~cm signal. Thus, this approach will open up new avenues for probing inflation with upcoming 21~cm surveys.

% Point 5: Objectives and findings
In this work, we employ MFs to investigate the signatures of 
primordial bump-like features on the 21~cm signal from the EoR. 
Our objectives are threefold: (i) to characterise how bump models 
modify the morphology of the density, spin temperature, neutral 
hydrogen, and brightness temperature fields; (ii) to determine 
whether MFs can distinguish bump models -- particularly at the 
``turnover scale'' where the global signal is insensitive -- from 
the fiducial model; and (iii) to examine whether primordial 
signatures can be distinguished from variations in astrophysical 
parameters governing the EoR.

%\sout{Hence we have simulated the 21 cm signal including this  redshift range and used morphological statistic - Minkowski Functionals to distinguish the bump model from fiducial power law form of inflation. We have also attempted to understand the physical effects of inflationary features on the minkowski functionals of 21 cm signal and related cosmological fields contributing to it. We have also tested different EoR scenarios against inflationary features of bump model to see potential degeneracies in the 21 cm signal. This is essential to check the possibility of constraining inflationary features using future SKA data. The method used is general and applicable on a wide class of inflationary features models.}
%
We modify the initial conditions of the semi-numerical simulation 
code \texttt{21cmFAST} v3~\cite{Mesinger:2011,Murray:2020trn} to 
incorporate bump-like features and compute MFs on the resulting 
three-dimensional fields. We find that MFs are highly sensitive 
to both the amplitude and location of primordial features, and 
can distinguish bump models from the fiducial case across a wide 
range of redshifts. Crucially, MFs can identify bump models at 
the turnover scale, where the global 21~cm signal of reionisation 
history is indistinguishable from the fiducial model. Furthermore, 
by comparing the morphological signatures of primordial features 
with those of EoR parameter variations, we identify redshift ranges 
where the two can be distinguished, demonstrating the potential of 
MFs for constraining inflationary physics with future SKA observations.

The paper is organized as follows. Section~\ref{sec:s2} presents 
overviews of the physics of inflationary particle production,
and the epoch of reionisation, 
along with the semi-numerical 
simulations used in this work. Section~\ref{sec:s3} presents the definitions of 
Minkowski functionals, along with a summary 
of their redshift evolution for the key
fields of the EoR for the case of a fiducial model. Section~\ref{sec:s4} presents our results 
on the morphological signatures of inflationary 
particle production for fixed EoR physics. Section~\ref{sec:s5} 
explores potential degeneracies between inflationary particle 
production and EoR physics. We conclude with a summary of our main 
results and discussion in section~\ref{sec:s6}.

%%%%%%%%%%%%%%%%%%%%%%%%%%%%%%%%%%%%%%%%%%%%%%%%%%%%%%%%%%%%%%%%%%%%%
\section{Physics of inflationary particle production and reionization}
\label{sec:s2}
%%%%%%%%%%%%%%%%%%%%%%%%%%%%%%%%%%%%%%%%%%%%%%%%%%%%%%%%%%%%%%%%%%%%%
In this section, we provide an overview of the inflationary particle production mechanism central to this work. We then review the physics governing cosmic dawn and the EoR, and describe our simulation framework for modeling the high-redshift 21~cm signal.

%%%%%%%%%%%%%%%%%%%%%%%%%%%%%%%%%%%%%%%%%%%%%%%%%%%%%%%%%%%%%%%%%%%%%
\subsection{Particle production during inflation}
\label{sec:s21}
%%%%%%%%%%%%%%%%%%%%%%%%%%%%%%%%%%%%%%%%%%%%%%%%%%%%%%%%%%%%%%%%%%%%%

We focus on a class of inflationary models involving bursts of particle production during inflation, which predict bump-like features in the primordial power spectrum~\cite{Chung:1999ve,Barnaby:2009mc,Barnaby:2009dd,Pearce:2017bdc}. Such particle production mechanisms are well motivated in inflation models based on higher-dimensional gauge theories~\cite{Furuuchi:2015foh,Furuuchi:2020klq,Furuuchi:2020ery}. The resulting power spectrum can be written approximately as~\cite{Pearce:2017bdc,Naik:2022mxn}
\begin{equation}
\mathcal{P}_\mathcal{R}(k)
=
A_s \left(\frac{k}{k_0}\right)^{n_s-1}
+
A_{\rm I} \sum_i \left(\frac{f_1(x_i)}{f_1^{\rm max}}\right)
+
A_{\rm II} \sum_i \left(\frac{f_2(x_i)}{f_2^{\rm max}}\right)\,,
\label{eq:bump}
\end{equation}
where $A_s$ is the amplitude of the scalar perturbations, $n_s$ is the scalar spectral index, $k_0 = 0.05\,{\rm Mpc}^{-1}$, is the pivot scale, and $A_{\rm I}$ and $A_{\rm II}$ are the amplitudes of the dominant and subdominant contributions from particle production, respectively. The scale dependence of these contributions is given by the dimensionless functions
\be
f_1(x_i) \equiv
\frac{\left[\sin(x_i)-\mathrm{Si}(x_i)\right]^2}{x_i^3}\,, \quad 
f_2(x_i) \equiv
\frac{-2x_i\cos(2x_i)+(1-x_i^2)\sin(2x_i)}{x_i^3}\,, \label{eq:f1f2}
\ee
where $x_i \equiv k/k_i$, and $\mathrm{Si}(x) \equiv \int_0^{x}\frac{\sin t}{t}\,dt$ is the sine integral. The functions $f_1(x)$ and $f_2(x)$ attain their maximum values at $f_1^{\rm max} \simeq 0.11$ and $f_2^{\rm max} \simeq 0.85$, respectively. The parameter $k_i\,({\rm Mpc}^{-1})$ specifies the location of the $i^{\rm th}$ feature in the primordial power spectrum. The peak of the $i^{\rm th}$ dominant contribution occurs at
\begin{equation}
    k_{{\rm peak},i} \simeq 3.35 \times k_i\,,
    \label{eq:kpeak}
\end{equation}
while the subdominant contribution peaks at $x_i \simeq 1.25$.
\begin{figure}[t]
   \centering
   \includegraphics[width=0.7\textwidth]{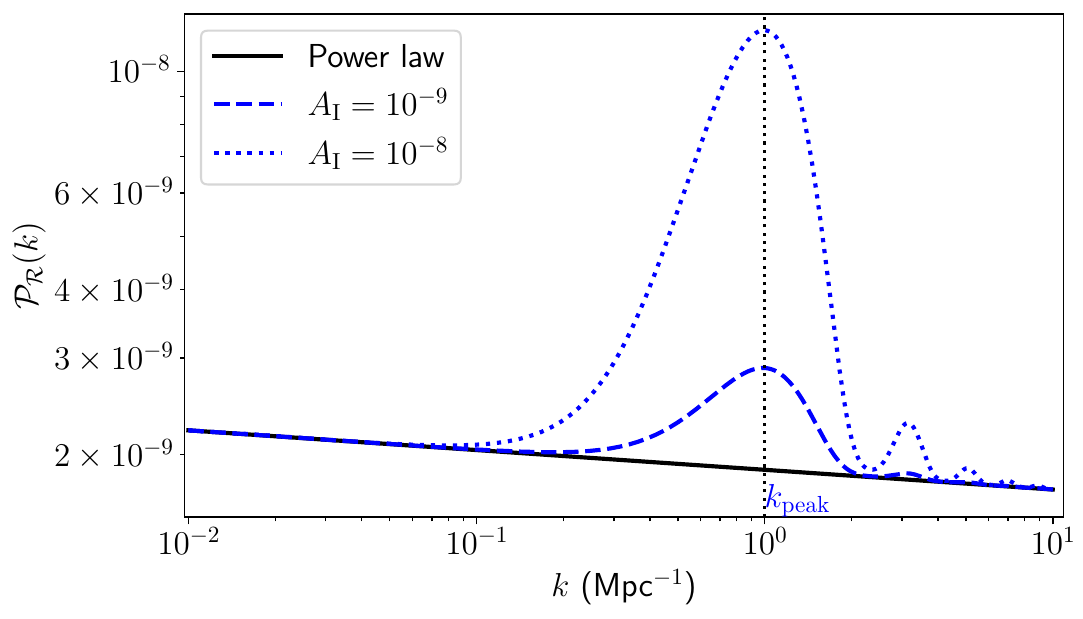}
   \caption{The primordial power spectrum with a bump-like primary feature at $k_{\rm peak} = 1\,{\rm Mpc}^{-1}$ arising from particle production during inflation.}
   \label{fig:bump_model}
\end{figure}
In this work, we consider a single burst of particle production during inflation, resulting in bump-like features parameterised by the amplitude $A_{\rm I}$\footnote{The subdominant amplitude is not an independent parameter and is related to $A_{\rm I}$ by $A_{\rm II} \simeq (2.9\times 10^{-6})\, A_{\rm I}^{5/7} \left[\ln A_{\rm I}^{4/7} + 24\right]$.} and the peak location $k_{\rm peak}\,({\rm Mpc}^{-1})$.

The signatures of bump-like features have been investigated using CMB data from the \textit{Planck} satellite~\cite{Naik:2022mxn}, which placed upper limits on their amplitudes over comoving wavenumbers $2\times 10^{-4} \lesssim k\,({\rm Mpc}^{-1}) \lesssim 0.15$. The impact of such features on the 21~cm signal has been examined through both the 21~cm power spectrum~\cite{Naik:2022wej} and the globally averaged signal~\cite{Naik:2025mba}, demonstrating the potential of upcoming 21~cm observations to constrain these models. While ongoing experiments primarily target measurements of the global signal and power spectrum, mapping the 21~cm brightness temperature field can provide complementary information. The 21~cm signal from cosmic dawn and the EoR is expected to be inherently non-Gaussian, and inflationary models predicting primordial features can introduce additional non-Gaussianity~\cite{Pearce:2017bdc}. It is therefore essential to employ multiple statistical measures to fully characterise the signal.

%In the next section, we provide a brief overview of the physics of the 21~cm signal. In section~\ref{sec:s3}, we outline the methodology used in this work to quantify the 21~cm signal and extract possible inflationary signatures. 

%Particle production during inflation also introduces non-Gaussianity, which was  quantified using the bispectrum in ref.~\cite{Pearce:2017bdc}. It was analytically calculated that the bispectrum also shows a localised bump-like feature when the three momenta are at the scale $k\sim k_{\rm peak}$, which left the horizon during the particle production.

%%%%%%%%%%%%%%%%%%%%%%%%%%%%%%%%%%%%%%%%%%%%%%%%%%%%%%%%%%%
\subsection{Physics and simulation of the 21~cm signal}
\label{sec:s22}
%%%%%%%%%%%%%%%%%%%%%%%%%%%%%%%%%%%%%%%%%%%%%%%%%%%%%%%%%%%%
The 21~cm signal from neutral hydrogen at a given redshift $z$ is observed along the line of sight against a background radio source, which at high redshifts is predominantly the CMB. The differential brightness temperature associated with the 21~cm signal is defined as the difference between the brightness temperature of the hydrogen cloud and the CMB temperature, and is given by~\cite{Furlanetto_2006}:
\begin{equation} \label{eq:brightness_temp}
\delta T_{b}(\vec{x},z) \ \approx
    27\, x_{\rm HI}\, 
    (1+ \delta_{\rm b}) \, 
    \left(\frac{H}{{dv_r/dr} + H}\right)
    \left(1 - \frac{T_\gamma}{T_S}\right)
    \left(\frac{1+z}{10} \frac{0.15}{\Omega_M h^2}\right)^{1/2}
    \left(\frac{\Omega_{b} h^2}{0.023}\right)
    {\rm mK}\,,
\end{equation}
where $x_{\rm HI}$ is the neutral hydrogen fraction, $\delta_{\rm b}$ is the baryon overdensity, $dv_r/dr$ is the comoving velocity gradient along the line of sight, $T_\gamma$ is the CMB temperature, and $T_S$ is the spin temperature, which characterizes the excitation of the 21~cm hyperfine transition. In subsequent parts of the paper we will refer to $\delta_b$ simply as `density field' as is commonly done. 
% Each of these cosmological fields contributes to the brightness temperature field $\delta T_b$.

%It is defined by ratio of number density of hydrogen atoms in two hyperfine states of 1S.\\
%\begin{equation}
%    \frac{n_1}{n_0} = \frac{g_1}{g_0}e^{-T_*/T_S}.
%\end{equation}
%$T_*$ = hc/k$\lambda_{21cm}$ = 0.068K.\\

%Move to simulations --> We use the following background cosmological parameters: $\Omega_\Lambda^0 = 0.72$, $\Omega_M^0 = 0.28$, $\Omega_b^0 = 0.046$, $H_0$ = 70 km $s^{-1}$ ${\rm Mpc^{-1}}$, $h$ = 0.7, $\rho_c^0$ = $\frac{2H_0^2}{8\pi G}$ =$0.92 \times 10^{-26}$ kg $m^{-3}$.\\

From eq.~\eqref{eq:brightness_temp}, we see that the detectability of the 21~cm signal depends critically on the spin temperature; the signal 
is observable only when it deviates from the CMB temperature.
% The spin temperature conveys information
% about astrophysical processes that contribute to it.
The spin temperature encodes information about the underlying astrophysical processes and is determined by three competing mechanisms: (i) absorption and emission of 21~cm photons from the radio background, primarily the CMB; (ii) collisions with other hydrogen atoms and with free electrons; and (iii) resonant scattering of Ly$\alpha$ photons, which induces spin flips via an intermediate excited state (the Wouthuysen--Field effect). Since the rates of these processes are fast compared to the Hubble time, the spin temperature can be well approximated by the equilibrium condition (see, e.g., \cite{Furlanetto_2006, Wouthuysen1958PIRE...46..240F, Field1952AJ.....57R..31W} for a detailed explanation):
\begin{equation} 
   T_S^{-1} = \frac{ T_\gamma ^{-1} + x_\alpha  T_\alpha ^{-1} + x_c  T_K ^{-1}}{1+x_\alpha +x_c}\,,
   \label{eq:Tspin}
\end{equation}
% $T_\alpha$ = color temperature of LY$\alpha$ radiation. \\
% $T_k$ = gas kinetic temperature by recoil in scattering\\
% $x_c$ = coupling coefficient for atomic collisions\\
% $x_\alpha$ = coupling coefficient for scattering of Ly$\alpha$ photons\\
% For more details refer to \cite{21cmfast};
where $T_\alpha$ is the color temperature of the Ly$\alpha$ radiation field, $T_K$ is the gas kinetic temperature 
($T_\alpha \approx T_K$ in most astrophysical conditions ), $x_c$ is the collisional coupling coefficient, and $x_\alpha$ is the Ly$\alpha$ coupling coefficient (Wouthuysen--Field coupling).

%\begin{equation}
%    x_c^i \equiv \frac{C_{10} T_*}{A_{10}T_\gamma}
%\end{equation}
% Here superscript $i$ denotes different colliding species. $C_{10}$ is collisional excitation rate, $A_{10}$ is einstein coefficient.
%\begin{equation}
%    x_\alpha = \frac{4P_\alpha}{27 A_{10}}\frac{T_*}{T_\gamma}
%\end{equation}
%Here $P_\alpha$ is scattering rate of Ly$\alpha$ photons.
%%%%%%%%%%%%%%%%%%%%%%%%%%%%%%%%%%%%%%%%%%%%%%%%%%%%%%%%%%%%%%%
\subsection*{Simulations of reionization}
%\label{sec:s23}

To model the cosmological fields entering eq.~\eqref{eq:brightness_temp}, we employ  \texttt{21cmFAST v3}\footnote{\url{https://github.com/21cmfast/21cmFAST}}~\cite{Mesinger:2011,Murray:2020trn}, a semi-numerical simulation tool widely used for generating cosmological 21~cm signals. The code produces three-dimensional realisations of the evolved density field $\delta_b$, neutral hydrogen fraction $x_{\rm HI}$, comoving peculiar velocity gradient $dv_r/dr$, and spin temperature $T_S$, which are then combined to compute the 21~cm brightness temperature $\delta T_b$ via eq.~\eqref{eq:brightness_temp}.

The brightness temperature is sensitive to various astrophysical processes, which are parameterised in \texttt{21cmFAST} as follows:
%(see, e.g., \cite{Greig:2017jdj} for details):
\begin{itemize}
    \item $T_{\rm vir}$\,(K): the minimum virial temperature of haloes hosting star-forming galaxies, equivalently characterised by the cut-off mass $M_{\rm min}$ below which star formation is negligible;
    \item $\zeta$: the ionising efficiency of high-redshift galaxies, which depends on the escape fraction of ionising photons $f_{\rm esc}$, its power-law scaling with halo mass $\alpha_{\rm esc}$, the star formation efficiency $f_\ast$, its power-law scaling with halo mass $\alpha_\ast$, the number of ionising photons per baryon $N_\gamma$, the typical number of recombinations $n_{\rm rec}$, and the star formation timescale $t_\ast$ (expressed as a fraction of the Hubble time);
    \item $L_{X<2\,{\rm keV}}/{\rm SFR}$: the soft-band X-ray luminosity (below 2\,keV) per unit star formation rate and the minimum energy of X-ray photons that can escape the host galaxy $E_0$.
\end{itemize}

\renewcommand{\arraystretch}{1.3} 
\begin{table}[t]
\centering
\resizebox{\textwidth}{!}{
\begin{tabular}{|lll|}
\hline
\texttt{\bf Parameters} &  \texttt{\bf Model}& \texttt{\bf Values/Range} \\ 
\hline
\multirow{3}{*}{Simulation} 
& & Box length $= 300$\,Mpc \\
& & Resolution $= 3$\,Mpc \\
& & Redshift range: $z = 5 - 35$ \\
& & Smoothing scale = 6\,Mpc\\
\hline
Background cosmology & & $\Omega_\Lambda = 0.72$, $\Omega_M = 0.28$, $\Omega_b = 0.046$, $H_0 = 70$\,km\,s$^{-1}$\,Mpc$^{-1}$ \\
\hline
\multirow{2}{*}{Primordial power spectrum} 
& Fiducial & $\ln (10^{10} A_s) = 3.044$, $n_s = 0.965$, $\sigma_8 = 0.811$ \\
& Bump models & $A_{\rm I} \in [10^{-9}, 10^{-8}]$, $k_{\rm peak} \in [0.4, 0.5, 0.6]$\,Mpc$^{-1}$ \\
\hline
\multirow{2}{*}{Astrophysical parameters} 
& Fiducial & $\zeta = 30$, $T_{\rm vir} = 4.69 \times 10^4$\,K, $\log(L_X/{\rm erg\,s^{-1}}) = 40.5$ \\
& EoR models & $\zeta \in [20, 45]$, $T_{\rm vir} \in [4.5, 5.0] \times 10^4$\,K, $\log(L_X/{\rm erg\,s^{-1}}) \in [38.5, 41]$ \\
\hline
\end{tabular}
}
\caption{Simulation configuration and model parameters used in this work.}
\label{tab:t1}
\end{table}

% We simulate 21~cm signal along with cosmological fields - $\delta_b$, $x_{HI}$ and $T_S$ using 21cmFAST in the redshift range of z = 35 to 5. We smooth the signal using gaussian kernel to 6 Mpc scale. We generate this signal for the fiducial cosmology model - power law inflation model and fixed EoR parameters $\zeta = 30$, $T_{\rm vir} = 4.69 \times 10^4 K$, $\log{L_x} = 40.5$ (table \ref{tab:t1}). We want to see signatures of bump model in 21~cm signal. For this we incorporate the bump inflationary power spectrum (figure \ref{fig:bump_model}) in initial conditions of the simulation and generate the cosmological fields for bump model.
We simulate the 21~cm signal along with the underlying 
cosmological fields ($\delta_b$, $x_{\rm HI}$, and $T_S$) 
over the redshift range $z = 5$ -- $35$. The resulting maps 
are smoothed with a Gaussian kernel to a resolution of 
6\,Mpc, and we calculate the morphological statistics of these 
smoothed fields. As our baseline, we consider a fiducial 
model without primordial features, adopting a power-law 
primordial power spectrum and fixing the astrophysical 
parameters to $\zeta = 30$, $T_{\rm vir} = 4.69 \times 10^4$\,K, 
and $\log(L_X/{\rm erg\,s^{-1}}) = 40.5$, as summarised in 
table~\ref{tab:t1}. To investigate signatures of the 
inflationary models with bump-like features 
in the 21~cm signal, we incorporate the corresponding primordial power spectrum 
(figure~\ref{fig:bump_model}) into the initial conditions 
and generate the cosmological fields. 
We refer to such models hereafter 
as ``bump models''. 

% Naik et al. in \cite{Naik:2025mba}
% studied the bump inflation model using the global signal, finding that it can discriminate different models. However, if the bump location corresponds to the  scale ${k_{\rm \bf p} \sim 0.5\,{\rm Mpc}^{-1}}$, the global signal is unable constrain, irrespective of redshift. Hence, we re-examine the bump inflation model about ${k_{\rm \bf p} \sim 0.5\,{\rm Mpc}^{-1}}$ using Minkowski functionals to provide additional constraining power. For this reason we are have chosen ${k_{\rm \bf p} = [0.4,0.5,0.6]\,{\rm Mpc}^{-1}}$ and $A_{\rm I} = [10^{-8},\ 10^{-9}]$ which is within bounds from \cite{Naik:2022mxn}.
The effects of bump models were recently studied in detail using the 
global 21~cm signal in ref.~\cite{Naik:2025mba}, which 
demonstrated the ability of global 21~cm profiles to 
discriminate between the effects of different values of amplitude 
$A_{\rm I}$ and location of the bump $k_{\rm peak}$. However, for primordial 
features located near a ``turnover scale'', $k_{\rm peak} \simeq k^{\rm turn}$, 
the global signal shows no effects of them regardless of 
redshift. We therefore examine bump models in this regime 
using Minkowski functionals, 
which can provide complementary constraining power. 
Specifically, we consider 
$k_{\rm peak} \in \{0.4, 0.5, 0.6\}\,{\rm Mpc}^{-1}$, which 
includes the turnover scale $k^{\rm turn} \simeq 0.5\,{\rm Mpc}^{-1}$ 
for the EoR parameters listed in Table~\ref{tab:t1}~\cite{Naik:2025mba}. 
For the amplitude, we consider $A_{\rm I} \in \{10^{-9}, 10^{-8}\}$, 
consistent with the upper limit $A_{\rm I} \lesssim 10^{-8}$ 
obtained from the global 21~cm signal using the optical depth 
to reionisation measured by \textit{Planck} 
(see figure~11 of Ref.~\cite{Naik:2025mba}).

%\purple{Expanding the RHS of eq \ref{eq:brightness_temp} and scaling $\delta T_b$ by spatially independent constant, 
%\begin{equation}
%    \frac{\delta T_{b}(\vec{x},z)}{C} = x_{HI}+x_{HI}\delta_b - \frac{x_{HI}\delta_b}{T_s}-\frac{x_{HI}\delta_b T_\gamma}{T_s}
%    \label{eq:Tb_exp}
%\end{equation}
%\purple{We take the global average of equation \ref{eq:Tb_exp} and plot the individual terms contributing to brightness temperature in figure \ref{fig:f2}. Note that we have neglected the $\del_r v_r$ term as it is of the order of $\sim 10^{-9}$. Different terms dominate at different redshifts as can be seen in figure \ref{fig:f2} - at early redshifts the contribution to $\delta T_b$ is mostly from $\delta_b$ and $T_s$ and $x_{HI}$ is 1 at almost all spatial points; so the resulting $\delta T_b$ is summation of second and third term. At intermediate redshifts $z \sim 10 \ \rm to \ 20 $ the contribution is mostly from the third term (green curve) as $x_{HI}$ and $T_s$ both contribute to $\delta T_b$. At late redshift $\delta T_b$ starts following $x_{HI}$ field as can be seen from red curve tending to the blue one for $z<10$.}
 
 \begin{figure}[tbp]
    \centering
    \includegraphics[width=\linewidth]{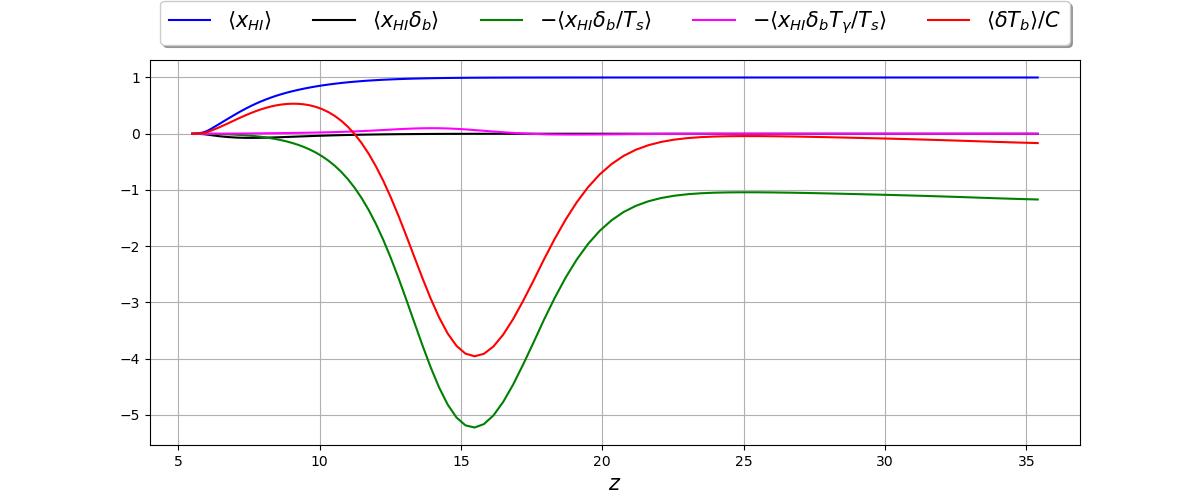}
    \caption{Redshift dependence of the mean values of different terms that contribute to the mean brightness temperature (see eq.~\eqref{eq:Tb_exp}). The spatially constant quantity $C$ is defined by eq.~\eqref{eq:Cdef}.} 
    % Brightness temperature field $\delta T_b$ in terms of the constituent field terms (refer equation \ref{eq:Tb_exp}) as function of redshift}
    \label{fig:f2}
\end{figure}
%%%%%%%
To understand the redshift-dependent physical contributions to 
$\delta T_b$, we expand the right-hand side of 
eq.~\eqref{eq:brightness_temp} as a sum of constituent 
terms and take the spatial average. This gives
\begin{equation}
 \frac{1}{C}\langle   \delta T_{b}(\vec{x},z)\rangle = \langle x_{\rm HI}\rangle + \langle x_{\rm HI}\delta_b\rangle - \bigg\langle\frac{x_{\rm HI}\delta_b}{T_S} \bigg\rangle -\bigg\langle \frac{x_{\rm HI}\delta_b T_\gamma}{T_S} \bigg\rangle, 
    \label{eq:Tb_exp}
\end{equation}
where the spatially constant prefactor $C$ is defined as
\begin{equation}
    C = 27 \left( \frac{1+z}{10} \frac{0.15}{\Omega_M h^2} \right)^{1/2} 
    \left( \frac{\Omega_b h^2}{0.023} \right).
    \label{eq:Cdef}
\end{equation}
The redshift evolution of each term on the right-hand side of 
eq.~\eqref{eq:Tb_exp} is plotted in figure~\ref{fig:f2} for 
the fiducial primordial and EoR model. Since the velocity gradient 
$dv_r/dr \ll H$, 
%is of order $\sim 10^{-9}$, 
we set the factor containing 
it to unity.
From the figure, we see that different terms dominate at different epochs. At early times 
($z > 20$), the contribution to $\delta T_b$ arises primarily 
from $\delta_b$ and $T_S$, while $x_{\rm HI} \approx 1$. At 
intermediate redshifts ($10 \lesssim z \lesssim 20$), the third 
term (green curve) dominates as both $x_{\rm HI}$ and $T_S$ 
contribute significantly. At lower redshifts ($z < 10$), 
$\delta T_b$ approximately traces $x_{\rm HI}$. Although this 
classification into high, intermediate, and low redshift regimes 
is approximate and model-dependent, we adopt it in subsequent 
sections to organise our results.

% We simulated these cosmological fields using 21cmFAST for the redshift range z $\sim$ 5 to 35. 
%\blue{We use a simulation box of length 300 Mpc with a resolution of 3 Mpc. We simulate the above-mentioned cosmological fields in the redshift range $z$ $\sim$ 35 to 5.
%Our simulations include the complete spin temperature calculations. }
% The simulation parameters used are as follows:\\
% \begin{enumerate}
% \item Box length = 300 Mpc,
% \item Resolution = 3 Mpc,
% \item Redshift range $z$ = 5 to 35,
% \item Redshift spacing = 1.
% \end{enumerate}
% At each redshift we have simulated $\delta_b$, $x_{\rm HI}$, Ts, $\partial_r v_r$, $\delta T_b$ field values at each pixel (size = 3Mpc) in $300\times300\times300$ Mpc 3D cube.\\
%\blue{We adopt the Friedmann-Lemaître-Robertson-Walker cosmological framework 
%with flat spatial geometry. 
%The concordance $\Lambda$CDM model
%is parameterized by 
%the baryon density 
%$\omega_b = \Omega_b h^2$, 
%the cold dark matter density 
%$\omega_{\text{cdm}} = \Omega_{\text{cdm}} h^2$,
%the present Hubble parameter $H_0$,
%the optical depth to reionization 
%$\tau$, 
%and $\sigma_8$, the variance of the density fluctuations 
%within	a sphere of $8 h^{-1}{\rm Mpc}$ radius.}

% \red{Explain the choice of the different parameter ranges.}

% \red{Mention smoothing here.}

% \red{Mention how particle production is incorporated. The power spectrum in \texttt{21cmFAST} is modified to include the bump 
% according to eq.~\eqref{eq:bump}.}

%%%%%%%%%%%%%%%%%%%%%%%%%%%%%%%%
\section{Minkowski Functionals and their behaviour for the fiducial model}  \label{sec:s3}

In this section, we review the formalism of Minkowski 
functionals for three-dimensional fields. We then describe 
the information that can be extracted by applying them to 
the 3D realisations of the fiducial model described in 
section~\ref{sec:s22}. This serves as a foundation for 
understanding the effects of inflationary signatures, 
which we examine in subsequent sections. 
%\sout{In the context of constraining EoR parameter space Glesser et al 2006 \cite{Gleser_2006} , Yoshiura et al 2016 \cite{10.1093/mnras/stw2701} and  Chen et al 2019 \cite{Chen_2019} have used 3D minkowski functionals.}
%\blue{***moving to the next subsection.}

%%%% Define excursion set and MFs
Given a smooth mean zero random field $f(\mathbf{x})$ with  standard deviation $\sigma_0$, the excursion set at threshold \(\nu\) is defined to be the set $Q(\nu) = \{ \mathbf{x} \, | \, f(\mathbf{x}) \geq \nu \sigma_0 \}$. 
$Q(\nu)$ contains all spatial regions where the field value exceeds a chosen threshold value. One can probe the morphology of the field across different levels by varying $\nu$. To quantify the geometry and topology of the excursion sets, we use the commonly used morphological statistics Minkowski Functionals (MFs). In three dimensions they are defined by the following volume averaged quantities,
%%%
\begin{equation}
			V_0 = \frac{1}{V} \int_{Q} dV \, ,	\quad V_1 = \frac{1}{6 V} \int_{\partial Q} \,dA \, , \quad    V_2 = \frac{1}{3 \pi V} \int_{\partial Q} G_2 \,dA \, , \quad    V_3 = \frac{1}{4 \pi^2 V} \int_{\partial Q} G_3 \,dA  \, ,
\end{equation}
%%%
where $\partial Q$ denotes the boundary surface defined by the condition $f = \nu\sigma_0$ and $V$ is the total volume occupied by the field.   For $V_2$ and $V_3$, the integrands are the mean curvature $G_2$, and the Gaussian curvature $G_3$ of the surface $\partial Q$. %  are $G_{2} = (R_{1}^{-1} + R_{2}^{-1})/2$ and $G_{3} = (R_{1}R_{2})^{-1}$, with $R_{1},R_{2}$ being the principle curvatures of the surface $\partial Q$. 
The four MFs describe the volume fraction ($V_0$), surface area density ($V_1$), integrated mean density ($V_2$) and integrated Gaussian curvature density ($V_3$) of $\partial Q$.

For a mildly non-Gaussian field, $V_i(\nu)$ takes the following simple closed form expressions as perturbative expansion in powers of $\sigma_0$, keeping upto first order \cite{Matsubara:2003,Hikage_2006}:
\begin{eqnarray}
    V_i(\nu) &=& \frac{1}{(2\pi)^\frac{i+1}{2}} \frac{\omega_3}{\omega_{3-i}\omega_i} \left(\frac{\sigma_1}{\sqrt{3}\sigma_0}\right)^i e^{-\nu^2/2} \bigg[ H_{i-1}(\nu) \nn\\ 
    && + \bigg( \frac{1}{6}S^{(0)}H_{i+2}(\nu)+\frac{i}{3}S^{(1)}H_i(\nu) + \frac{i(i-1)}{6}S^{(2)}H_{i-2}(\nu) \bigg)\s_0 + O(\sigma_0^2)\bigg],
    \label{eq:mf}
\end{eqnarray}
where $i=0,\ldots,3$, $\sigma_1=\la |\nabla f|^2\ra$, $H_n(\nu)$ are the probabilist Hermite polynomials, $H_{-1}=\sqrt{\frac{\pi}{2}}e^{\frac{\nu^2}{2}}{\rm erfc}\left(\frac{\nu}{\sqrt{2}}\right)  $, and %$\omega_n = \pi^{n/2}/\Gamma(n/2 +1)$; so that 
$\omega_0=1$, $\omega_1=2$, $\omega_2=\pi$ and $\omega_3 = \frac{4\pi}{3}$. 
% \red{make sure all factors are correct.}     
The quantities $S^{(j)}$ are the generalized skewness parameters defined by: 
\be
S^{(0)} \equiv \frac{\langle f^3 \rangle}{\sigma_0^4},\quad 
S^{(1)} \equiv -\frac{3}{4} \frac{\langle f^2 (\nabla^2 f)\rangle}{\sigma_0^2 \sigma_1^2} ,\quad S^{(2)} \equiv -\frac{9}{4} \frac{\langle (\nabla f).(\nabla f) (\nabla^2 f)\rangle}{\sigma_1^4}.
\label{eq:Si}
\ee
For Gaussian fields, eq.~\eqref{eq:mf} further simplifies because terms of order $\sigma_0$ and higher are zero.

Here the random field $f(\vec{x})$ can be one of the cosmological fields - density field ($\delta_b(\vec{x},z)$), neutral hydrogen fraction ($x_{\rm HI}(\vec{x},z)$), spin temperature ($T_S(\vec{x},z)$), brightness temperature fluctuation ($\delta T_b(\vec{x},z)$). These fields have spatial fluctuations as well as redshift dependence. The variation of the amplitudes and shapes of the MFs with redshift encode how different physical effects affect the fields, and can be used to distinguish different primordial and EoR models. In general, the fields are non-Gaussian and the shapes of the MFs will depart from the Gaussian expectations. We will see in the next subsection that $\delta_b$ is mildly non-Gaussian at the redshifts and smoothing scales studied here, and their MFs can be described by eq.~\eqref{eq:mf}. In contrast, the other fields are strongly non-Gaussian, for which perturbative expansions of the MFs about the Gaussian expectations are not applicable.   

\subsection{Morphology of EoR fields for the fiducial model}
\label{sec:s32}
%%%%%%%%%%%%%%%%%%%%%%%%%%%%%%

We first examine the morphology of the EoR fields for 
the fiducial model, which adopts a power-law primordial 
power spectrum without the bump-like feature and the 
astrophysical parameters listed in table~\ref{tab:t1}. 
With this parameter set, reionization reaches 50\% at 
$z = 7.57$ and completes by $z \sim 6$. The simulated 
fields from \texttt{21cmFAST} are smoothed with a Gaussian 
filter to a resolution of 6\,Mpc, and we compute the MFs 
on the smoothed fields at different threshold values $\nu$ using the publicly available code \texttt{PyMIN}\footnote{\url{https://github.com/dkn16/pyMIN.}} 
\cite{Diao_2024}. Since the EoR fields do not have zero mean, in general, we use threshold values of the normalised field $\tilde f=\frac{f-\la f\ra}{\sigma_0}$. 
The features in the MFs of the brightness temperature 
encode information about structure formation and reionization 
through their dependence on the density field $\delta_b$, 
spin temperature $T_S$ and neutral fraction $x_{\rm HI}$.

\begin{figure}[tbp]
{\centering
(a)\\
\includegraphics[width=1\linewidth]{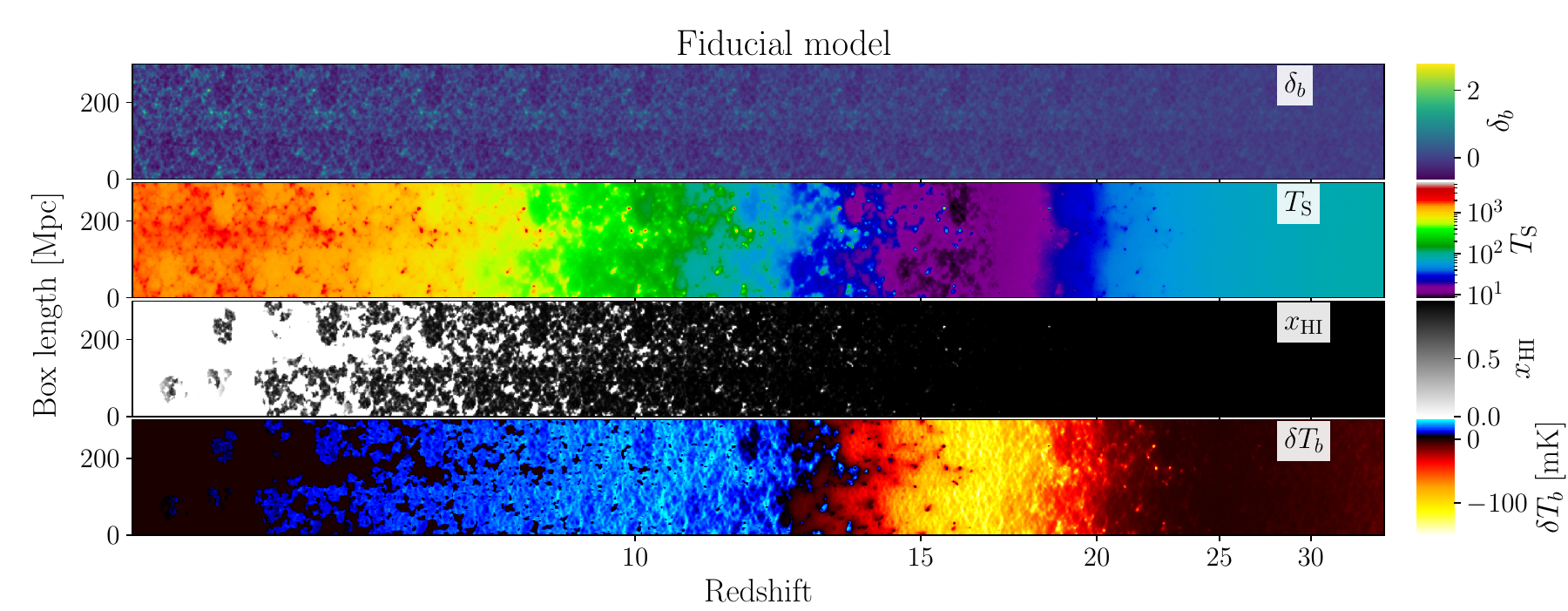}\\
(b)\\
\hskip -.5cm 
\includegraphics[width=0.7\linewidth]{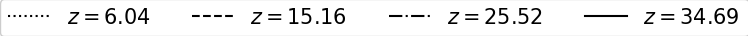}\\}
\includegraphics[width=0.99\linewidth]{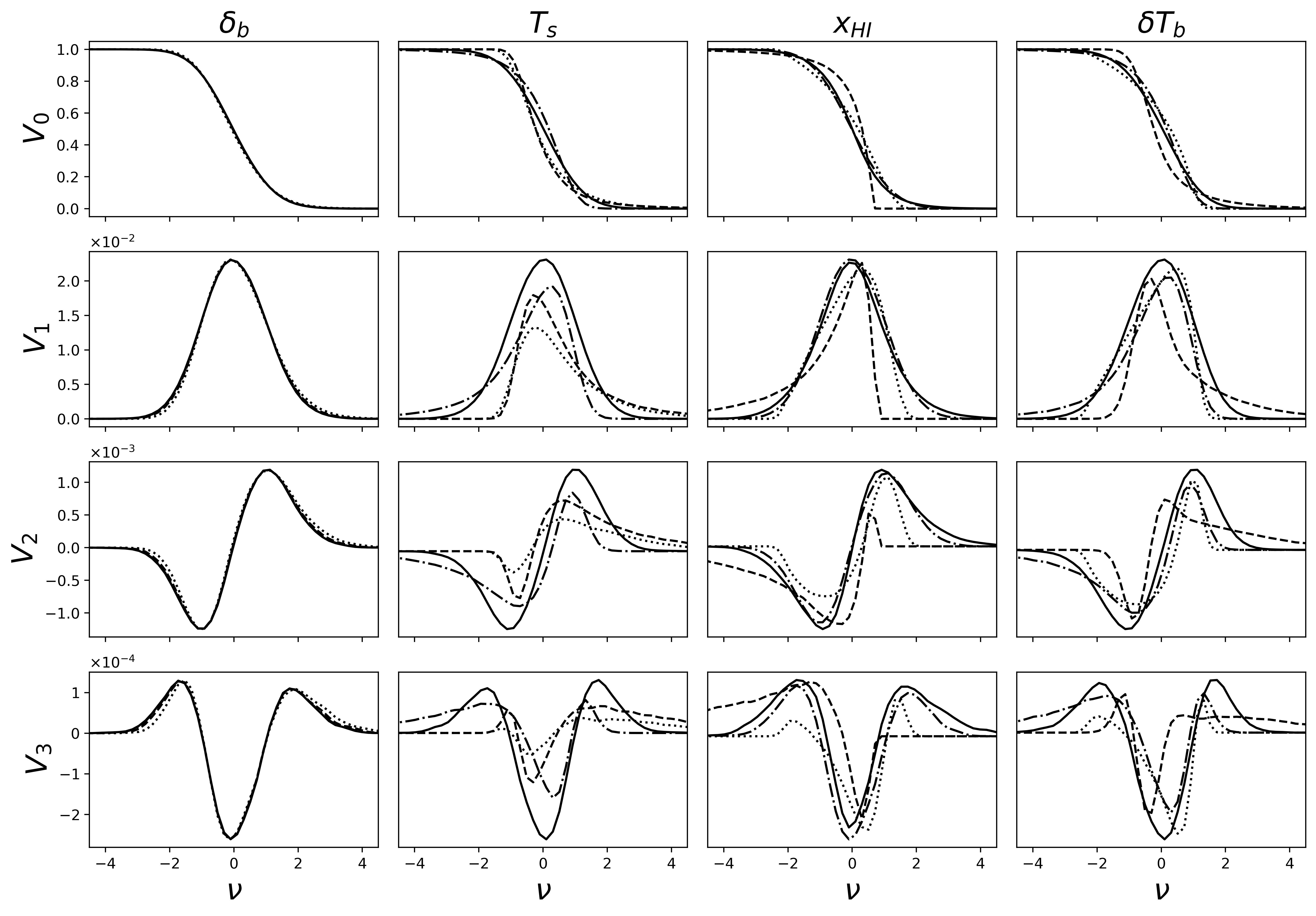}
\caption{(a) Lightcones of $\delta_b$, $T_S$, $x_{\rm HI}$, and $\delta T_b$ (rows) for the fiducial model; colours indicate the field values as shown by the respective colorbars. (b) Minkowski functionals $V_0$, $V_1$, $V_2$, and $V_3$ (rows) as functions of threshold $\nu$ for the same fields as in panel (a) (columns), shown at selected redshifts indicated by different line styles in the legend.}  
\label{fig:fid}
\end{figure}

% \red{(e.g., cite e.g. Gleser:2006, Yoshiura:2016, Chen:2019)}.
Minkowski functionals for fields on three-dimensional space have previously been used to investigate different EoR  models
and to constrain the astrophysical parameter space
\cite{Gleser_2006,yoshiura2016,Diao_2024,Chen_2019}.
Here, we elaborate on the morphology of each physical field that contributes to  $\delta T_{b}$, namely, $\delta_b,\, T_{S}$ and $x_{\rm HI}$,  
to understand how different quantities affect the morphology 
of the brightness temperature. Figure~\ref{fig:fid}(a) shows 
the lightcones for these fields. The Minkowski functionals 
for the 3D coeval cubes - $V_0$, $V_1$, $V_2$, and $V_3$ - are 
plotted in the rows of figure~\ref{fig:fid}(b) at selected 
redshifts corresponding to key evolutionary stages 
(early structure formation, IGM heating, and the 
progression of reionization). We discuss each of these 
in detail below.
%%%%%%%%
%In the {\it fiducial model}, the astrophysical parameters are fixed 
%according to \cite{Park:2018ljd}:
%$M_{\rm min} = 5\times 10^8 M_\odot$ (or $\log T_{\rm vir} = 4.69897$),
%$\zeta = 30$ (i.e., $f_{\rm esc} = 0.1$, $f_\ast = 0.05$, and 
%$N_\gamma = 5000$),
%$\alpha_\ast = 0.5$,
%$\alpha_{\rm esc} = -0.5$,
%$t_\ast = 0.5$,
%$E_0 = 0.5$~keV, and 
%$L_{X<2{\rm {keV}}}/$SFR $ = 10^{40.5}~\text{erg s}^{-1} M_\odot^{-1} {\rm yr}$.
%The cosmological parameters are set following the 
%best-fit values from the \textit{Planck} 2018 results \cite{Akrami:2018odb}:
%$\Omega_b h^2 = 0.022$,
%$\Omega_{\text{cdm}} h^2 = 0.120$,
%$h = 0.6736$,
%$\tau = 0.054$,
%$\sigma_8 = 0.811$, and
%$n_s = 0.965$. 
%%%%%%%%%
%We quantify the 3D simulated fields of the fiducial model using the Minkowski functionals defined in the previous section. 
%%%%%%%

%%%%%%%%%%

\vskip .2cm
\noindent{\bf Density field:} The density field and its 
MFs for the fiducial model are shown in the first row of 
panel (a) and first column of panel (b) of figure~\ref{fig:fid}. 
The distribution of density field is nearly Gaussian 
at high redshift ($z \sim 35$), as the initial density 
fluctuations after recombination are assumed to be described 
by a Gaussian random field. Accordingly, the MFs closely 
follow the analytical expressions for a mildly non-Gaussian field given 
in equations~\eqref{eq:mf}. Deviations from Gaussianity are 
expected to increase with time due to the nonlinear growth 
of structures driven by gravitational clustering at lower 
redshifts. Although the redshift dependence of the MFs of 
the density field is mild, a slight increase in asymmetry 
of the MF curves from $z = 35$ (solid line) to $z = 6$ 
(dotted line) indicates the growing deviation from 
Gaussianity over time (see discussion of the skewness parameters shown in figure~\ref{fig:sigma_evo} in section~\ref{sec:bump_density}.) 
Consequently, the contribution of the density field to the 
redshift evolution of the brightness temperature morphology 
is expected to be minimal.
    
\vskip 2mm  
\noindent{\bf Spin temperature field:} The MFs for $T_S$ 
are shown in the second column of figure~\ref{fig:fid}(b). 
At $z = 35$, the MFs of $T_S$ closely follow those of 
$\delta_b$. This is because, at such high redshifts, there 
are no Ly$\alpha$ sources, and spin temperature fluctuations 
arise solely from collisional coupling ($x_c$), which is 
proportional to the baryon density (cf.\ eq.~\eqref{eq:Tspin}). 
At $z < 30$, Ly$\alpha$ coupling becomes significant and 
$T_S$ becomes highly non-Gaussian, with deviations more 
pronounced than in $\delta_b$ across all four MFs. At 
$z \sim 25$, $T_S$ couples to the gas kinetic temperature 
$T_K$, which is colder than the CMB ($T_K < T_\gamma$). 
Near overdense regions, fluctuations in both density and 
Ly$\alpha$ flux enhance this coupling, causing a sharp 
drop in $T_S$. This manifests as longer tails at negative 
thresholds in $V_1$, $V_2$, and $V_3$. Conversely, at 
$z \lesssim 15$, heating from the first sources 
produces localised increases in $T_S$ around overdense 
regions, leading to longer tails at positive thresholds.
In addition to the change in the profile of the 
MF curves, their amplitudes also decrease with decreasing 
redshift, reflecting the reduced contrast in $T_S$ 
fluctuations as heating progresses. 
Given this strong sensitivity to astrophysical processes, 
$T_S$ is expected to be one of the dominant contributors to the 
morphology of the brightness temperature.

\vskip 2mm
\noindent{\bf Neutral hydrogen field:} As shown in the 
third column of figure~\ref{fig:fid}(b), the MFs of the 
neutral hydrogen fraction $x_{\rm HI}$ follow those of 
$\delta_b$ at high redshifts, similar to $T_S$. This is 
expected since neutral hydrogen was the most abundant 
baryon species during this epoch, before ionising sources 
had formed. The field becomes non-Gaussian and negatively 
correlated with $\delta_b$ once the first sources begin 
ionising the surrounding medium, as seen in the curves 
for $z \lesssim 25$. At $z \sim 15$, the IGM is still 
nearly neutral (see the lightcone in panel (a) of 
figure~\ref{fig:fid}), with only a few locally ionised 
regions as heating has just begun. This results in longer 
tails at negative thresholds with a sharp drop at higher 
$\nu$. As $V_2$ characterises the connectivity of structures, 
it shows complex behaviour at lower redshifts due to 
percolation of ionised regions. By $z \sim 6$, the neutral 
fraction drops below 5\%, with the IGM dominated by ionised 
regions; consequently, $V_1$, $V_2$, and $V_3$ decrease at 
negative thresholds. The peaks at positive thresholds 
correspond to the few remaining isolated neutral hydrogen 
islands.
The strong evolution in $x_{\rm HI}$ morphology at lower redshifts directly 
drives the reionization signature in the brightness temperature field.

\vskip 2mm
\noindent{\bf Brightness temperature field:} 
As given in eq.~\eqref{eq:brightness_temp}, all the aforementioned 
fields contribute to the 21~cm signal. The MFs of $\delta T_b$ 
are shown in the last column of figure~\ref{fig:fid}(b). At 
$z \gtrsim 30$, fluctuations in $\delta T_b$ follow those in 
$\delta_b$. During the heating epoch ($15 \lesssim z \lesssim 25$), 
while most hydrogen remains neutral, $\delta T_b$ is primarily 
driven by $T_S$. Once reionization commences ($z \lesssim 15$), 
$x_{\rm HI}$ fluctuations dominate; by $z \sim 6$ (dotted line), 
the MFs of $\delta T_b$ closely follow those of $x_{\rm HI}$. 
Thus, the MFs of $\delta T_b$ trace the morphology of different 
contributing fields at different epochs. Having established 
this fiducial behaviour, we now examine how primordial features 
modify these MF signatures.

%%%%%%%%%%%%%%%%%%%%%%%%%%%%%%%%%%%%%%%%%%%%%%%%%%%%%%%%%%%%%%%%%%
\section{Results - effects of particle production during inflation }
\label{sec:s4}
%%%%%%%%%%%%%%%%%%%%%%%%%%%%%%%%%%%%%%%%%%%%%%%%%%%%%%

In this section, we present our results on the signatures of the bump model, characterised by the amplitude $A_{\rm I}$ and the primary peak location $k_{\rm peak}$ (see figure~\ref{fig:bump_model}).
The power spectrum in \texttt{21cmFAST} is modified to include this feature 
according to eq.~\eqref{eq:bump}. 

% \red{shall we mention here that we are taking only first term of eq 2.1???}

% To isolate the effects 
% of bump-like primordial features on the MFs of the 21~cm 
% fields, we fix the astrophysical parameters to the fiducial 
% values listed in Table~\ref{tab:t1}. We compute the MFs for 
% bump models with $A_{\rm I} \in \{10^{-9}, 10^{-8}\}$ and 
% $k_{\rm peak} \in \{0.4, 0.5, 0.6\}\,{\rm Mpc}^{-1}$, 
% and examine their effects at various redshifts 
% relative to the fiducial model discussed in the previous section.

% We show our results for each field as the excess MF relative to the fiducial model, defined as} 
% We show our results for each field's MF $V_i$, $i=\{0, 1, 2, 3\}$ for bump models and fiducial model. For the bump models we use parameter combinations of $k_{\rm peak}=0.4, \, 0.5,\, 0.6$, and $A_{\rm I}= 10^{-9}, 10^{-8}$ to calculate MFs as function of threshold($\nu$) in range $\nu = -5\sigma \ to \ +5\sigma$ where $\sigma$ is the variance of the field at a particular redshift $z$. 
% \be
% \Delta V_i = V_i^{\rm bump} - V_i^{\rm fid}\,,
% \label{eq:diffV}
% \ee
% where $i=\{0, 1, 2, 3\}$, 
% $V_i^{\rm bump}$ and $V_i^{\rm fid}$ denote the MFs calculated for bump models and the fiducial model, respectively. We calculate (\eqref{eq:diffV}) for different bump model parameter combinations $k_{\rm peak}=0.4, \, 0.5,\, 0.6$, and $A_{\rm I}= 10^{-9}, 10^{-8}$ as a function of threshold($\nu$) in range $\nu = -5\sigma \ to \ +5\sigma$ where $\sigma$ is the variance of the field at a particular redshift $z$.

%%%%%%%%%%%%%%%%%%%%%%%%%%%%%%%%%%%%%%%%%%%%%%%%%%%%%%%%%%%%%
\subsection{Background: Imprints of bump models on the 21~cm signal}
\label{sec:s41}
%%%%%%%%%%%%%%%%%%%%%%%%%%%%%%%%%%%%%%%%%%%%%%%%%%%%%%%%%%%%%

Before presenting our results on the morphology of fields 
for bump models, we briefly summarise their signatures on 
the 21~cm brightness temperature (see~\cite{Naik:2022wej,Naik:2025mba} 
for details).
%%%%%%%%
\begin{figure}[tbp]
    \centering
    (a) Density field\\
      \includegraphics[width=\linewidth]%, height=5cm]
      {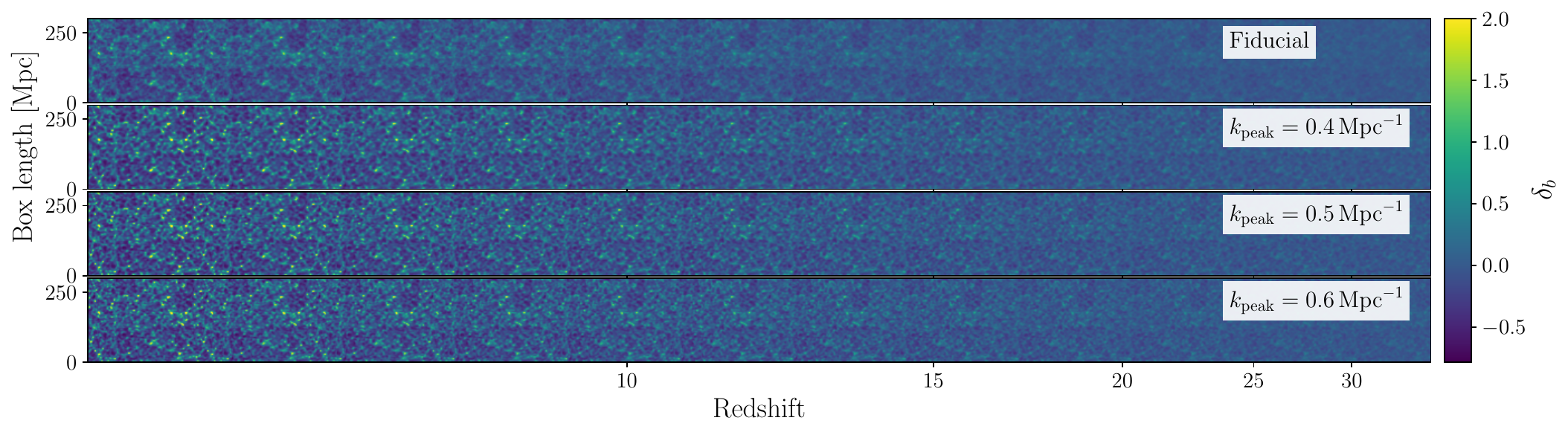}\\
      (b) Spin temperature \\
        \includegraphics[width=\linewidth]%, height=5cm]
        {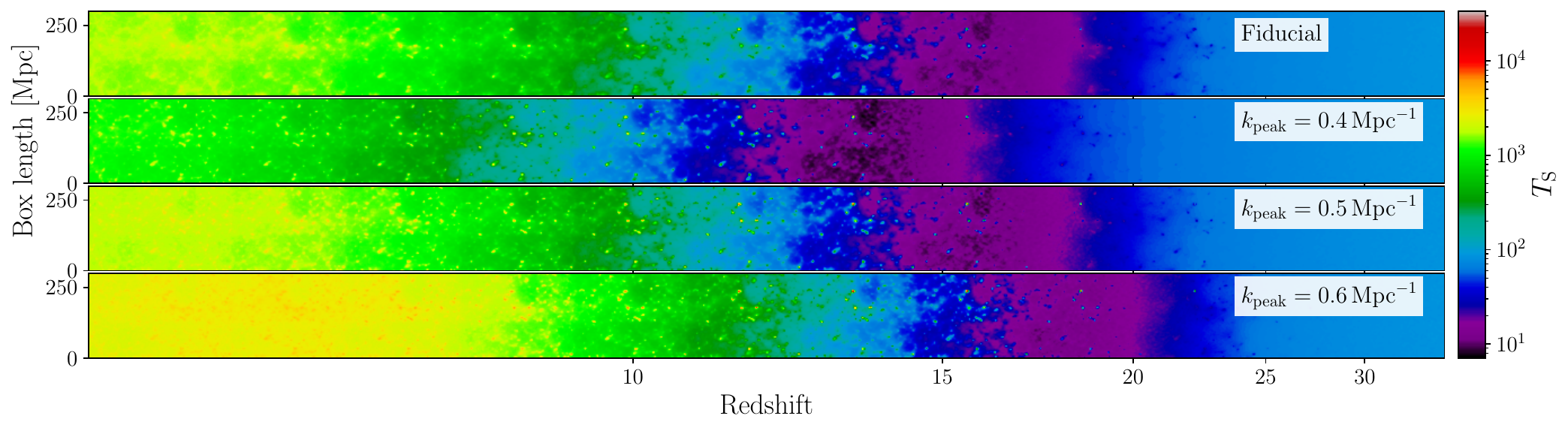}\\
        (c) Neutral fraction of hydrogen \\
        \includegraphics[width=\linewidth]%, height=5cm]
        {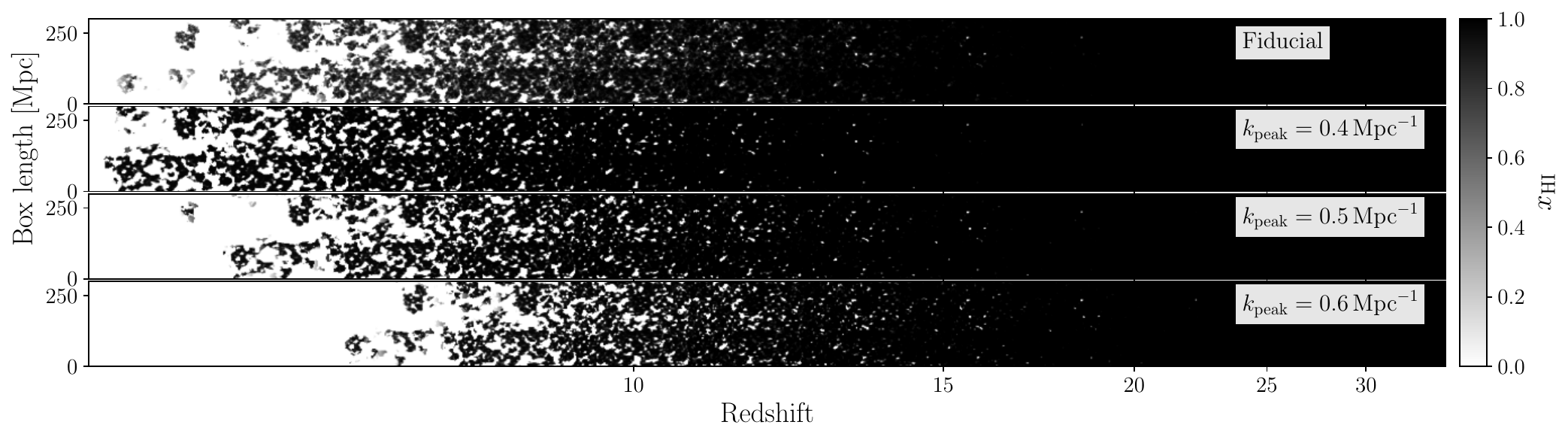}\\
        (d) Brightness temperature \\
        \includegraphics[width=\linewidth]%, height=5cm]
        {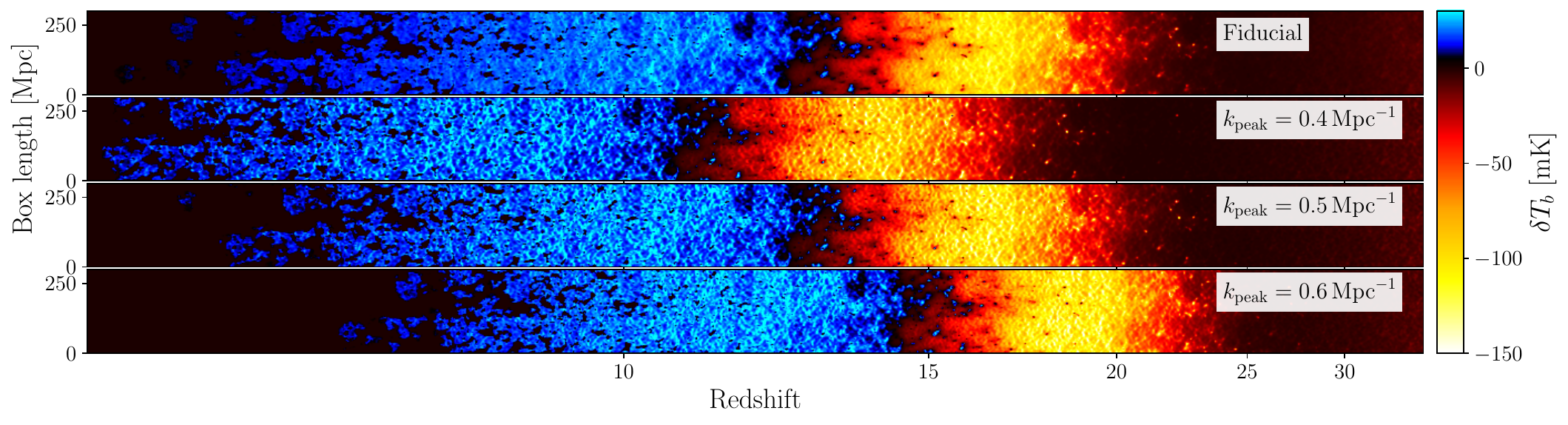}
    \caption{The simulated lightcones of all the fields for the fiducial and bump models.}
    %\mg{Mpc$^{-1}$ units for $k_{\rm peak}$ are dropped.}
    %\red{The kpeak legends are hard to see for $x_{HI}$. How about using a bright colour? Increase label sizes.}}
    \label{fig:lc_bump}
\end{figure}
%%%%

Eq.~\eqref{eq:brightness_temp} informs that the 21~cm fluctuations of the brightness temperature are inherited from the fluctuations of $\delta_b$, 
$T_S$, and $x_{\rm HI}$. In \texttt{21cmFAST}, the real-space 
density field is computed via an inverse Fourier transform of
\begin{equation}
    \delta_b(\vec{k}) = 
    \sqrt{\frac{V P(k)}{2}} (a_k + i b_k)\,,
\end{equation}
where $V$ is the box volume, $P(k)$ is the matter power spectrum which includes the bump-like 
features, and $a_k$, $b_k$ are Gaussian random 
variables with variance one~\cite{Mesinger:2007pd}. Primordial features thus 
directly affect $\delta T_b$ through $\delta_b$: a feature 
at low (high) $k_{\rm peak}$ enhances overdensities on large (small) spatial scales.

Furthermore, it was demonstrated in~\cite{Naik:2025mba} that 
primordial features significantly affect $x_{\rm HI}(\vec{x}, z)$ 
and $T_S(\vec{x}, z)$. The value of $k_{\rm peak}$ plays a 
crucial role in determining the ionisation history: higher 
(lower) values of $k_{\rm peak}$ lead to faster (slower) 
structure formation and correspondingly earlier (later) 
completion of reionization. 
Increasing the amplitude of the feature enhances such effects. 
Remarkably, at a critical 
``turnover'' scale $k_{\rm peak} = k^{\rm turn}$, bump-like 
features have negligible impact on the average reionization history
and consequently on the globally averaged 21~cm signal. The 
lightcones illustrating the evolution of different fields 
for the fiducial and bump models are shown in 
figure~\ref{fig:lc_bump}; for the EoR parameters considered 
here, $k^{\rm turn} = 0.5\,{\rm Mpc}^{-1}$. The temporal 
evolution of the neutral hydrogen field in panel (c) clearly 
demonstrates how bump models modify structure formation and 
alter the reionization history relative to the fiducial model.
This behaviour arises from the way primordial features modulate 
the collapsed fraction of haloes given by  the halo 
mass function (HMF). 
Due to the power spectrum normalization at 8 $h^{-1}{\rm Mpc}$, 
the enhancement of power at different scales introduced by the 
primordial features significantly alters the number density of halos.
Features with low (high) $k_{\rm peak}$ 
amplify overdensities on large (small) scales, thereby 
suppressing (enhancing) the number density of low-mass haloes 
and resulting in slower (faster) structure formation
(see 
appendix~\ref{app:hmf} for a summary of HMF modifications 
discussed in ref.~\cite{Naik:2025mba}). The analysis in 
ref.~\cite{Naik:2025mba} further revealed that $k^{\rm turn}$ 
depends on the value of $T_{\rm vir}$ since it sets the minimum halo mass for 
star formation, which in turn determines which scales dominate 
the ionisation budget.

With this understanding of how primordial features affect the 
21~cm signal, our objectives are threefold: (i) to extract 
signatures of bump models directly from the morphology of 
the fields; (ii) to investigate whether bump models at 
$k_{\rm peak} = k^{\rm turn}$, which remain undetectable in 
the global 21~cm signal, can be distinguished from the fiducial 
model through MFs; and (iii) to examine distinctive signatures 
of primordial features compared to those of astrophysical 
processes during the EoR. We address objectives (i) and (ii) 
in this section and defer (iii) to the next section. To isolate 
the effects of primordial features on the MFs, we fix all EoR 
parameters to their fiducial values (table~\ref{tab:t1}). We 
examine bump-like features with $k_{\rm peak}$ at and near 
the turnover scale, computing MFs for $k_{\rm peak}\,[{\rm Mpc}^{-1}] 
\in \{0.4, 0.5, 0.6\}$ and $A_{\rm I} \in \{10^{-9}, 10^{-8}\}$.

%%%%%%%%%%%%%%%%%%%%%%%%%%%%%%%%%%%%%%%%%%%%%%%%%%%%%%%%%%%%
\subsubsection{MFs of the density field}
\label{sec:bump_density}
%%%%%%%%%%%%%%%%%%%%%%%%%%%%%%%%%%%%%%%%%%%%%%%%%%%%%%%%%%%%
%
%%%%%%
\begin{figure}[tbp]
    \centering
    \includegraphics[width=0.9\linewidth]{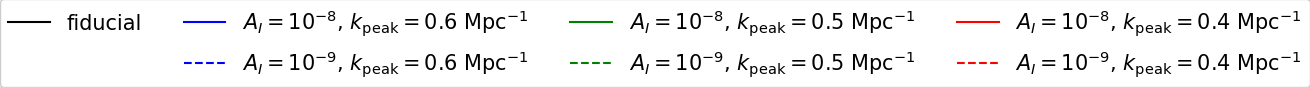}
    \includegraphics[width=\linewidth]{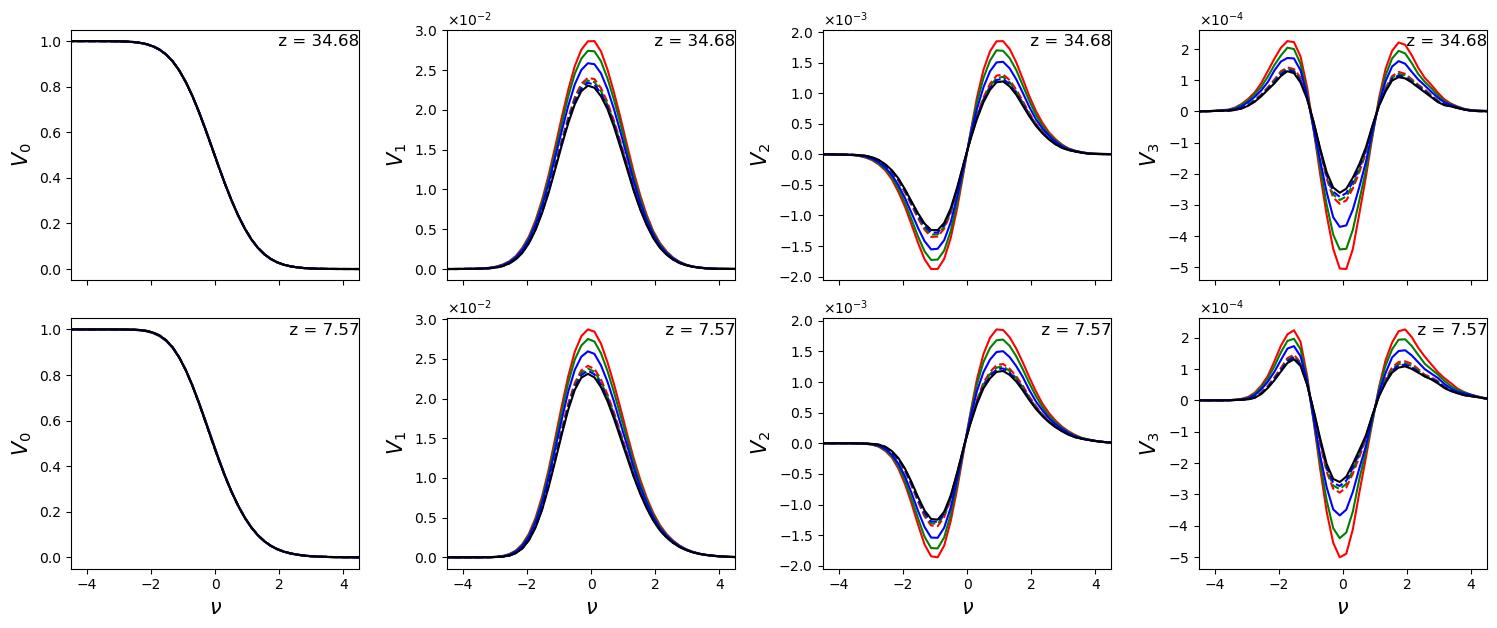}
    \caption{The Minkowski functionals %(as defined in eq.~\eqref{eq:diffV})
    for the density field $\delta_b(\vec{x})$ for models involving primordial bump-like features.
    The upper and lower panels show $ V_i$ at redshifts $z = 34.68$ and $z=7.57$, respectively.
    The fiducial model is indicated by solid black line. 
    %The values of $k_{\rm peak}$ are denoted by different colors and the values of $A_{\rm I}$ are denoted by different linestyles. 
    % \red{In legends of fig 4, 6, 7, 8 it is better to show fiducial on the left. If $\times 10^{-2}$ for  $V_1$, $\times 10^{-3}$ for  $V_2$, and $\times 10^{-4}$ for  $V_3$ are marked at the top left of each panel, some horizontal space can be saved and used to make the panels slightly bigger. Also, x-y- and z labels can be made larger.\\
    % Since we are not using the dots in the legends lets remove them.}
    %\red{ssn: remove fiducial from the legend here; make the yscale of $\Delta V_0$ common for both the redshifts. Change $k_{\rm peak}$ 
    %to $k_{\rm peak}$. Set tex font true.}
    }
    \label{fig:density_MF}
\end{figure}
\begin{figure}[tbp]
    %\hspace*{-3cm}
    %\includegraphics[width=0.9\linewidth]{figures/legend.png}
    \centering
    \includegraphics[width=0.87\linewidth]{figures/legend.png}\\
    \includegraphics[width=.9\linewidth]{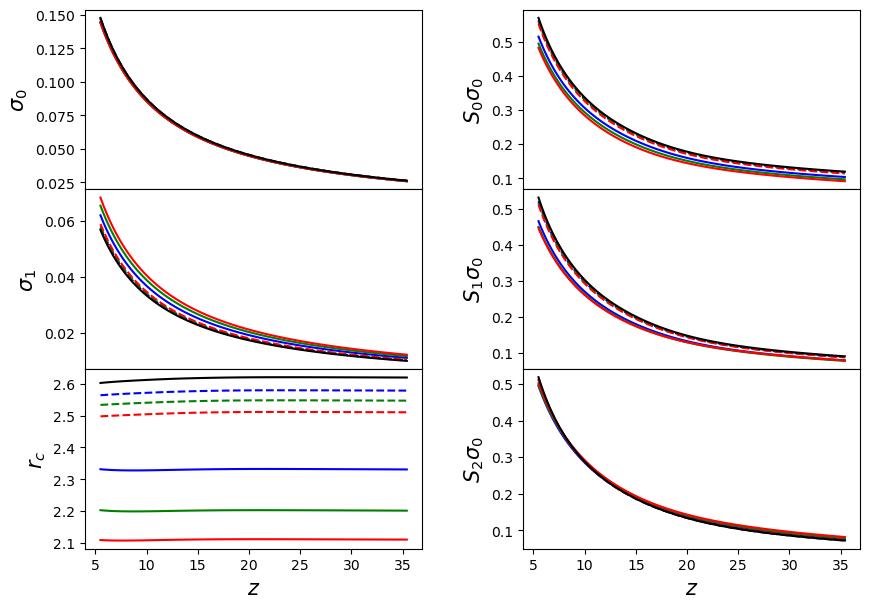}
    \caption{{\it Left}: redshift evolution of $\sigma_0$, $\sigma_1$ and $r_c$ for density field $\delta_b$ for the fiducial and bump models. {\it Right}: redshift evolution of $S^{(i)}\sigma_0$ for the same fields. } %The colour and linestyles is the same as in Figure~\ref{fig:density_MF}.} \red{How about the legends? Do you think the previous sentence suffices?} } %\red{Modify to show $\s_0, \s_1, r_c$ on left column and $S_0\s_0,\ S_1\s_0,\ S_2\s_0$ on right column. }}
    \label{fig:sigma_evo}
\end{figure}
%
%\begin{figure}[h] 
%    \centering
%    \includegraphics[width=\linewidth]{figures/den_sigmas_rc12.jpg}
%    \includegraphics[width=\linewidth]{figures/den_sigma0_sm2.jpg}
%    \caption{The redshift evolution of $\sigma_0$, $\sigma_1$ and $r_c=\sigma_0/\sigma_1$ and the non-gaussianity parameters for density field.
%    \blue{The fiducial model is indicated by solid black line. 
%    The values of $k_{\rm peak}$ are denoted by different colors and the values of $A_{\rm I}$ 
%    are denoted by different linestyles.}
%    \red{ssn: the cyan line at zero may not necessary. Change $k_{\rm peak} $ to $ k_{\rm peak}$ in the legend. }
%    } %\blue{keep only density}}
%    \label{fig:sigma_evo}
%\end{figure}

The lightcones for the density fields of the bump models are 
shown alongside the fiducial model in panel (a) of 
figure~\ref{fig:lc_bump}. Although the effects of different 
$k_{\rm peak}$ values are barely discernible by visual 
inspection, quantitative analysis using MFs reveals clear 
differences in morphology. In figure~\ref{fig:density_MF}, 
we present the MFs $V_k$ at redshifts $z = 34.68$ and 
$z = 7.57$, spanning early and late epochs, respectively\footnote{Since 
the MFs of the density field exhibit minimal redshift 
dependence, we present results at only two representative 
redshifts.}.
% We identify three 
% key findings: (a) the temporal evolution in $V_i$ is negligible; 
% (b) decreasing $k_{\rm peak}$ 
% enhances the deviation from the fiducial case; and 
% (c) increasing the amplitude $A_{\rm I}$ systematically enhances 
% the differences relative to the fiducial model, consistent 
% across all MFs and redshifts. We provide 
% physical interpretations of these behaviours below.

As in the fiducial case, the profiles of MFs of the density field 
do not change significantly with redshift. However, bump models 
enhance the MF amplitudes relative to the fiducial model. This 
enhancement arises from additional morphological complexity 
introduced by primordial features, and is amplified by increasing 
$A_{\rm I}$. To understand this behaviour, we examine the 
quantities controlling the MF amplitudes. Since $\delta_b$ 
exhibits mild non-Gaussianity within our redshift range, the 
MF amplitudes are primarily determined by $\sigma_0$ (the field 
variance) and $\sigma_1$ (the variance of the field gradient). 
The ratio $r_c \equiv \sigma_0/\sigma_1$ defines a characteristic 
length scale of structures in the density field, with MF 
amplitudes scaling inversely as powers of $r_c$. The left column of figure~\ref{fig:sigma_evo}  shows that both $\s_0$ and $\s_1$ grow with $z$ at roughly the same rate, and hence  $r_c$ remains nearly independent of redshift, 
explaining the negligible temporal evolution in the MFs. The right column shows the three skewness parameters $S^{(i)}\s_0$ defined by eq.~\eqref{eq:Si} (we show with the $\s_0$ factor since this gives the size of the non-Gaussian corrections, upto the Hermite polynomials and the constant prefactors). They grow with redshift as the density field becomes increasingly more non-Gaussian due to gravitational clustering. For the smoothing scale of 6\,Mpc employed here we see that the density field is remains mildly non-Gaussian till $z\sim 6$, as indicated by the fact that $S^{(i)}\s_0$ remains below one.

The bump model with $k_{\rm peak} = 0.4\,{\rm Mpc}^{-1}$ exhibits 
the maximum deviation from the fiducial model due to its lowest 
$r_c$ value. 
% As discussed previously, lower $k_{\rm peak}$ 
% amplifies large-scale overdensities. 
At our chosen smoothing 
scale, $\sigma_0$ values 
for all models coincide with the fiducial case, while $\sigma_1$ 
scales inversely with $k_{\rm peak}$. Consequently, 
$r_c \propto k_{\rm peak}$, leading to $V_k \propto 1/k_{\rm peak}$. 
Thus, bump models with lower $k_{\rm peak}$ enhance large-scale 
overdensities, thereby increasing MF amplitudes.
Conversely, for sufficiently high $k_{\rm peak}$, the 
introduced small-scale overdensities are suppressed by 
the smoothing, resulting in negligible effects 
on the MFs.
In summary, the MFs of $\delta_b$ are primarily sensitive 
to $k_{\rm peak}$ through its effect on the characteristic 
scale $r_c$ of density structures. %\red{The higher values of $S^{(i)}\s_0$ for the bump models compared to the fiducial one indicate that the bump models induce additional non-Gaussianty.} 

\subsubsection{MFs of the spin temperature field}\label{sec:bump_spintemp}

% At early redshift spin temperature field follows the density fluctuations because the only contribution to $T_S$ is from the collisional term which is proportional to the number density of colliding species in the IGM. This can also be seen in minkowski functionals difference plot of Ts in figure \ref{fig:ts_MF} being similar to the figure \ref{fig:density_MF} at redshift z = 34.7.\\
\begin{figure}[t]
    \centering
    \includegraphics[width=0.9\linewidth]{figures/legend.png}
    \includegraphics[width=\linewidth]{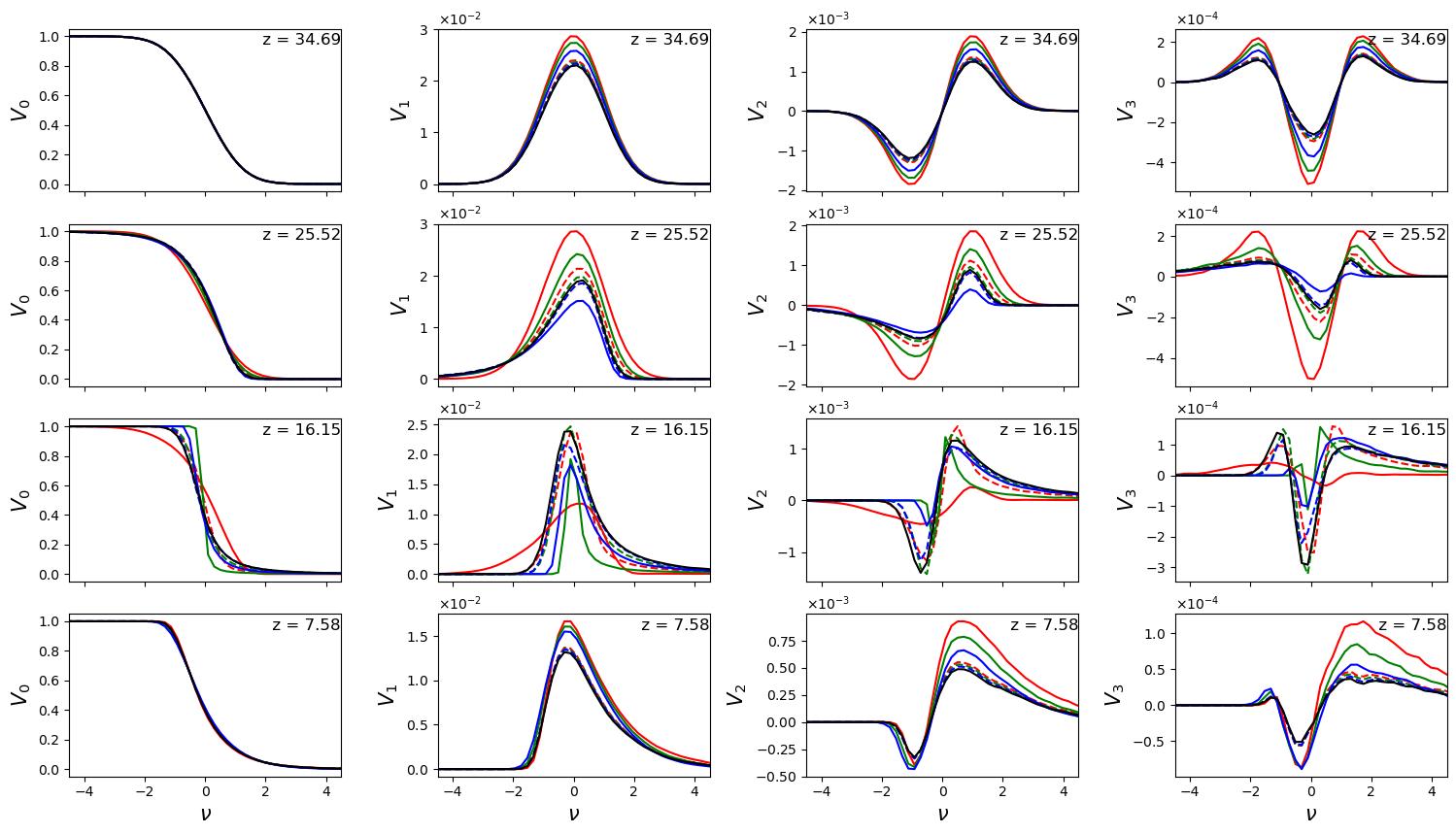}
    \caption{The Minkowski functionals 
    for the spin temperature field $T_S(\vec{x})$ for models involving primordial bump-like features.
    The fiducial model is indicated by solid black line. 
    The values of $k_{\rm peak}$ are denoted by different colors and the values of $A_{\rm I}$ 
    are denoted by different linestyles.}
    \label{fig:ts_MF}
\end{figure}
%%%%%%%
The lightcones for $T_S$ of the bump models are shown alongside the fiducial model in panel (b) of 
figure~\ref{fig:lc_bump}. One can observe strong variation with different 
$k_{\rm peak}$ values.  
The corresponding MFs  are shown in figure~\ref{fig:ts_MF}.  
The behaviour varies significantly across redshift regimes:
\begin{itemize}
    \item \textbf{$z \sim 35$:} The spin temperature field 
    closely follows the density field due to the absence of 
    heating and ionisation, as in the fiducial case.
    The MF amplitudes scale inversely with $k_{\rm peak}$, as discussed for 
    the density field.
    
    \item \textbf{$z \sim 25$:} In the fiducial model, the MFs 
    exhibit extended tails toward negative thresholds. This is  due to the 
    drop in $T_S$ around overdense regions which lead to an increase of the number of holes \cite{Kapahtia_2019}. The bump models also follow 
    this trend, with a special effect of $k_{\rm peak}$. 
    As $k_{\rm peak}$ increases, the MFs become increasingly asymmetric 
    because faster structure formation results in more non-Gaussian  
    characteristics in the $T_S$ field.
    For $k_{\rm peak} = 0.6\,{\rm Mpc}^{-1}$, 
    $T_S$ drops within more localised regions, resulting in 
    a sharper drop at positive thresholds, as  seen in the 
    $T_S$ map in figure~\ref{fig:lc_bump}(b).
    
    \item \textbf{$z \sim 16$:} In the fiducial model, heating 
    begins to affect the spin temperature, resulting in extended 
    tails toward positive thresholds. 
    This behaviour is also observed for $k_{\rm peak} = 0.5$ and 
    $0.6\,{\rm Mpc}^{-1}$. However, for $k_{\rm peak} = 0.4\,{\rm Mpc}^{-1}$, 
    slower structure formation delays the onset of heating, and 
    the $T_S$ field has not yet risen sufficiently at this redshift. 
    This results in distinctive MF behaviour compared to the other 
    models.
    
    \item \textbf{$z \sim 7.5$:} At this stage, reionization reaches $\sim$50\% 
    in the fiducial model. The lightcones show that $T_S$ increases 
    with higher $k_{\rm peak}$, and consequently the MF curves 
    exhibit extended tails toward positive thresholds.
\end{itemize}
In summary, the MFs of $T_S$ are sensitive to $k_{\rm peak}$ 
through its effect on the timing of structure formation and 
heating.

%\red{**********************}
%%%%%%%%%%%%%%%%%%%%%%%%%%%%%%%%%%%%%%%%%%%%%%%%%%%%%%%%%%%%
\subsubsection{MFs of the neutral hydrogen field}
\label{sec:bump_xHI}
%%%%%%%%%%%%%%%%%%%%%%%%%%%%%%%%%%%%%%%%%%%%%%%%%%%%%%%%%%%%
%\red{SSN: How bump models affect xHI field will be written here with the help of HMF and MFs}
\begin{figure}[t]
    \centering
    \includegraphics[width=0.9\linewidth]{figures/legend.png}
    \includegraphics[width=\linewidth]{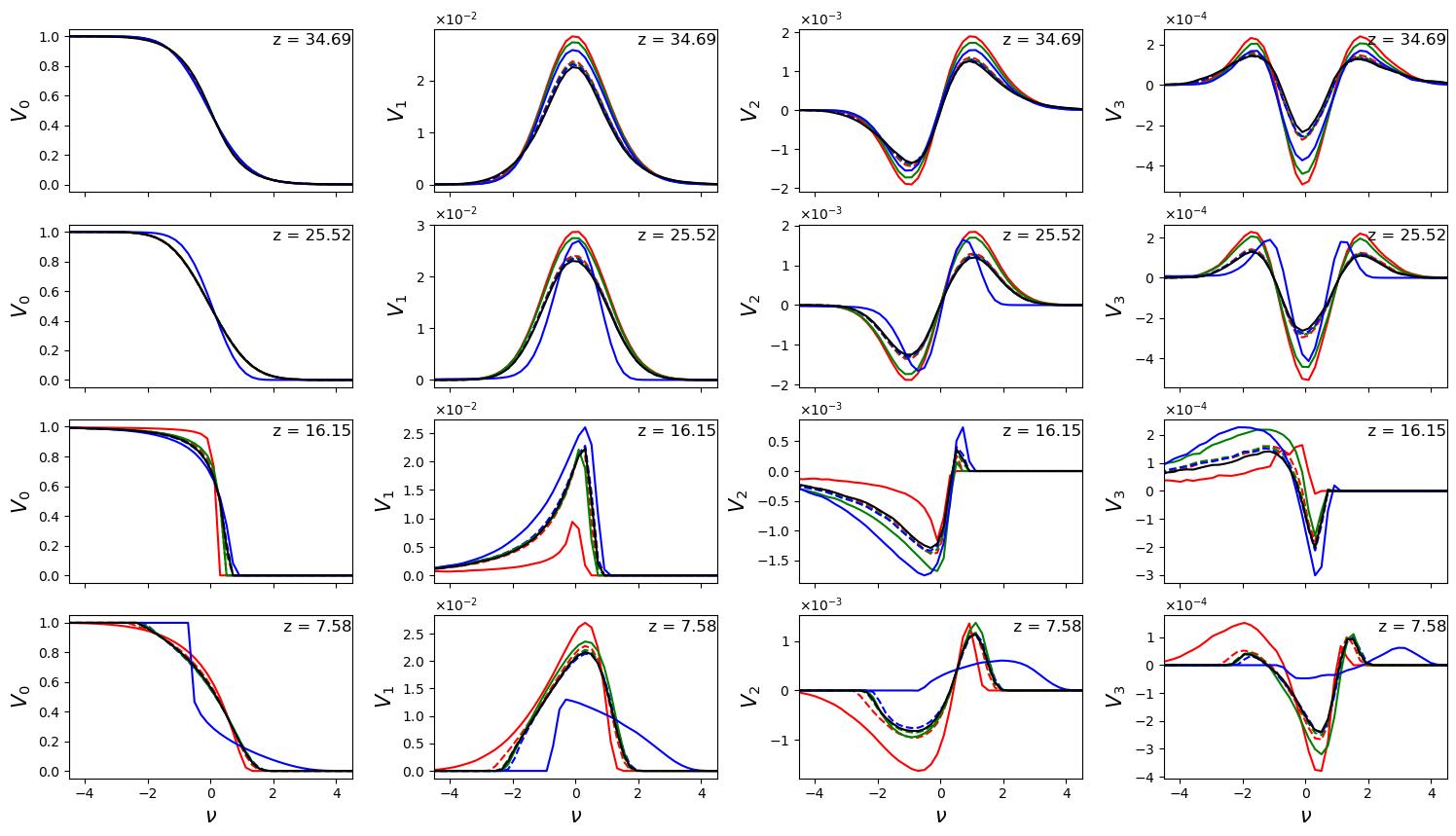}
    \caption{The Minkowski functionals 
    for the neutral fraction of hydrogen  $x_{\rm HI}(\vec{x})$ for models involving primordial bump-like features.
    The fiducial model is indicated by solid black line. 
    The values of $k_{\rm peak}$ are denoted by different colors and the values of $A_{\rm I}$ 
    are denoted by different linestyles.}
    \label{fig:xhi_MF}
\end{figure}

The lightcones for $x_{\rm HI}$ of all the models shown in panel (c) of 
figure~\ref{fig:lc_bump} again exhibit  strong variation with different 
$k_{\rm peak}$ values.  The corresponding MFs  are presented in 
figure~\ref{fig:xhi_MF}. The behaviour varies across redshift 
regimes:
\begin{itemize}
    \item \textbf{$z \sim 35$:} The neutral hydrogen field is 
    nearly Gaussian and follows the density field, as in the 
    fiducial case. The MF amplitudes scale inversely with 
    $k_{\rm peak}$, as discussed for the density field.
    
    \item \textbf{$z \sim 25$:} The fields remain close to Gaussian for 
    all models except $k_{\rm peak} = 0.6\,{\rm Mpc}^{-1}$, which 
    shows distinct departures from Gaussianity due to earlier 
    formation of small-scale structures and low-mass haloes.
    
    \item \textbf{$z \sim 16$:} Reionization proceeds faster for 
    higher $k_{\rm peak}$, producing more ionised regions. 
    Consequently, $k_{\rm peak} = 0.6\,{\rm Mpc}^{-1}$ exhibits 
    the largest MF amplitudes, while 
    $k_{\rm peak} = 0.4\,{\rm Mpc}^{-1}$ displays the lowest.
    
    \item \textbf{$z \sim 7.5$:} At this redshift, which 
    corresponds to 50\% reionization in the fiducial model, 
    the ionised fractions for $k_{\rm peak} = 0.4$, $0.5$, and 
    $0.6\,{\rm Mpc}^{-1}$ are 20\%, 47\%, and 86\%, respectively. 
    For $k_{\rm peak} = 0.6\,{\rm Mpc}^{-1}$, only scattered 
    neutral islands remain, producing a peak at positive 
    thresholds (most prominent in $V_3$). In contrast, for 
    $k_{\rm peak} = 0.4\,{\rm Mpc}^{-1}$, the peak at positive 
    thresholds arises from partially ionised regions as 
    reionization progresses more slowly.
\end{itemize}
In summary, the MFs of $x_{\rm HI}$ are sensitive to 
$k_{\rm peak}$ through its effect on the timing and morphology 
of reionization.
   
% %The MF difference plots of $V_1$, $V_2$ and $V_3$ for $x_{\rm HI}$ field is similar to density fluctuation field at early redshift z = 34.7. 
% The non-gaussianity parameters $S_0$, $S_1$ and $S_2$ (third row in figure \ref{fig:non-gaussianity}) also show that the field was mildly gaussian at early redshifts following the density field. At late redshifts $z < 25$, the $x_{\rm HI}$ field becomes highly non-gaussian. 
% %as can be seen from the plots of figure \ref{fig:xhi_nongaussianity}.
% At z = 7.57 when on average half of the neutral hydrogen atoms are ionized \blue{for the fiducial model}, the non Gaussianity parameters reduce again to fractional values.
% % In figure \ref{fig:xhi_nongaussianity} the red curve of bump model $k_{\rm peak}=0.4$ has higher $S_0$,$S_1$ and $S_2$ values at earlier redshift z = 25 then the blue curve of $k_{\rm peak} = 0.6$ model which shows the same trend but at z = 16. This indicates that ionization is happening earlier for higher $k_{\rm peak}$ bump model. This is consistent with Naik et. al. 2025 paper where they found same trend through global average of $\langle x_{\rm HI} \rangle$

%\red{**********************}

%%%%%%%%%%%%%%%%%%%%%%%%%%%%%%%%%%%%%%%%%%%%%%%%%%%%%%%%%%%%
\subsubsection{MFs of the brightness temperature field}
\label{sec:bump_dTb}
%%%%%%%%%%%%%%%%%%%%%%%%%%%%%%%%%%%%%%%%%%%%%%%%%%%%%%%%%%%%

\begin{figure}[t]
    \centering
    \includegraphics[width=0.9\linewidth]{figures/legend.png}
    \includegraphics[width=\linewidth]{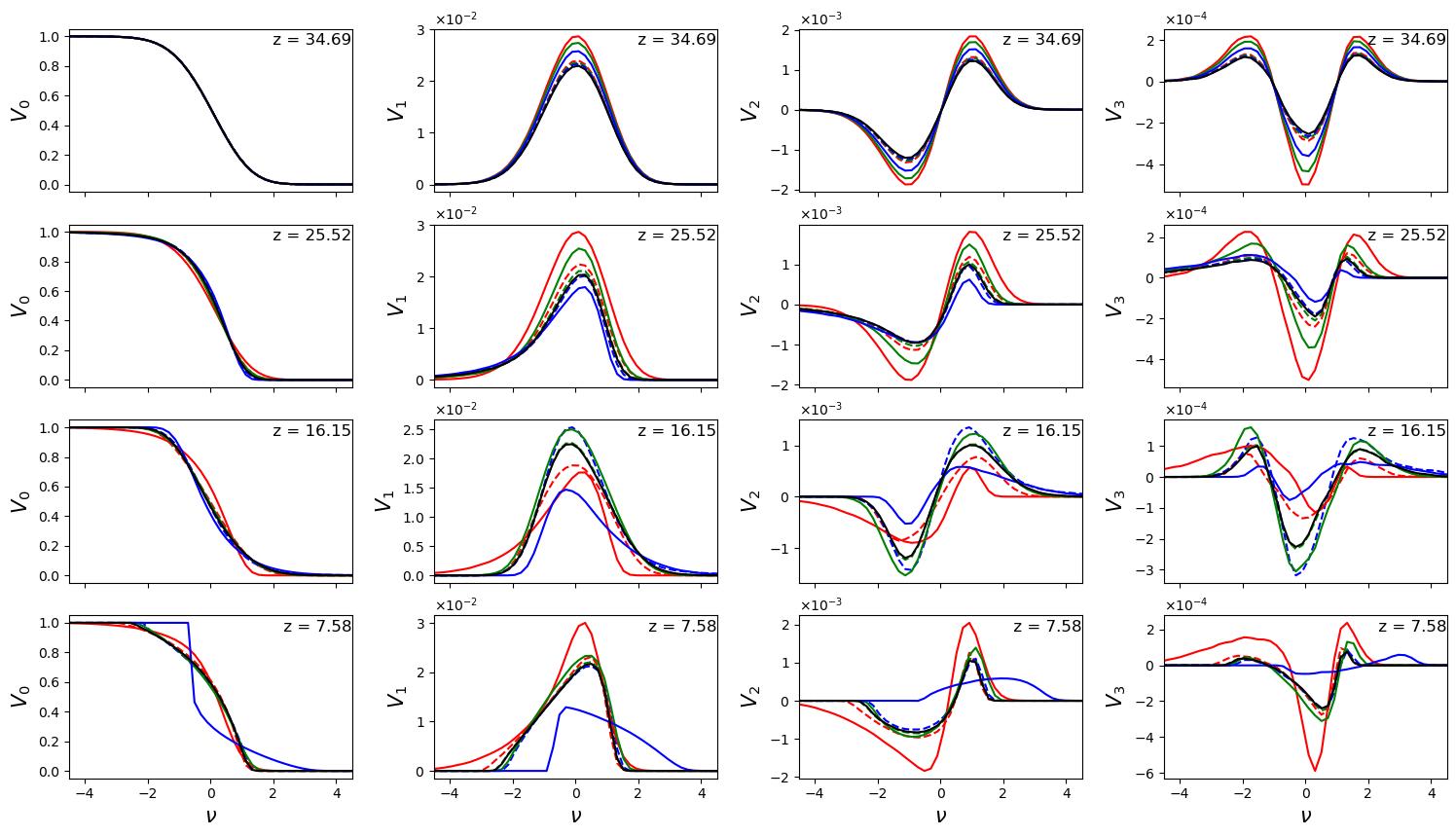}
    \caption{The Minkowski functionals 
    for the brightness temperature of the 21~cm signal, $\delta T_b(\vec{x})$, 
    for models involving primordial bump-like features.
    The fiducial model is indicated by solid black line. 
    The values of $k_{\rm peak}$ are denoted by different colours and the values of $A_{\rm I}$ 
    are denoted by different linestyles.}
    \label{fig:tb_MF}
\end{figure}

With insights from the preceding analysis of individual fields, 
we now examine the combined impact of bump models on the 21~cm 
signal. The lightcones depicting these effects are shown in 
figure~\ref{fig:lc_bump}(d). Visual inspection reveals significant 
morphological differences between bump models and the fiducial 
case, highlighting the potential for studying inflationary 
signatures in upcoming 21~cm brightness temperature maps.

We present the MFs for bump models along with the fiducial 
model in figure~\ref{fig:tb_MF}. At different evolutionary 
stages, fluctuations in $\delta T_b$ are dominated by different 
component fields as seen in figure \ref{fig:f2} and also reflected in both the lightcones and MFs:
\begin{itemize}
    \item \textbf{$z \sim 35$:} At this redshift, $\delta T_b$ 
    follows the density field $\delta_b$ (see figure~\ref{fig:density_MF}). 
    As discussed previously, fluctuations in $\delta_b$ are 
    modulated by structures of different scales introduced through 
    primordial features. For example, 
    $k_{\rm peak} = 0.4\,{\rm Mpc}^{-1}$, which enhances large-scale 
    structures, exhibits higher MF amplitudes compared to the 
    other models.
    
    \item \textbf{$z \sim 25$:} As $T_S$ begins coupling with 
    $T_K$, fluctuations in $\delta T_b$ follow those of $T_S$. 
    Analogous to the analysis in section~\ref{sec:bump_spintemp}, 
    all models except $k_{\rm peak} = 0.4\,{\rm Mpc}^{-1}$ have 
    begun departing from the almost Gaussian nature of the field. The 
    contributions from $\delta_b$ and $x_{\rm HI}$ fluctuations 
    are subdominant at this stage.
    
    \item \textbf{$z \sim 16$:} Around this redshift, the 
    brightness temperature fluctuations are influenced by all 
    three component fields. The extended tail toward negative 
    thresholds for $k_{\rm peak} = 0.4\,{\rm Mpc}^{-1}$ reflects 
    the combined morphological influence of both $T_S$ and 
    $x_{\rm HI}$. In the brightness temperature field, most 
    regions exhibit absorption of the 21~cm signal, resulting 
    in lower MF amplitudes at positive thresholds. For 
    $k_{\rm peak} = 0.6\,{\rm Mpc}^{-1}$, the extended tail 
    toward positive thresholds arises from localised emission 
    regions due to relatively faster heating and reionization 
    processes.
    
    \item \textbf{$z \sim 7.5$:} During the later stages of 
    reionization, $x_{\rm HI}$ provides the dominant contribution 
    to $\delta T_b$, as the 21~cm signal persists only at the 
    neutral hydrogen islands that remain after ionised bubbles 
    have merged to remove most neutral hydrogen from the universe. 
    This effect is evident from both the lightcones and the MFs 
    shown in the bottom rows of figures~\ref{fig:xhi_MF} and 
    \ref{fig:tb_MF}.
\end{itemize}

To summarise, the MFs of all fields -- $\delta_b$, $T_S$, $x_{\rm HI}$, 
and $\delta T_b$ -- exhibit distinct signatures of bump-like primordial 
features. Lower $k_{\rm peak}$ enhances large-scale structures and 
increases MF amplitudes, while higher $k_{\rm peak}$ accelerates 
structure formation and reionization. 
%The amplitude $A_{\rm I}$ primarily scales the magnitude of deviations from the fiducial model without qualitatively altering these trends. 
The amplitude $A_{\rm I}$ primarily scales the magnitude of deviations from the fiducial model. %\sout{ without qualitatively altering these trends}.} 
However, the higher value $A_{\rm I}=10^{-8}$ does introduce additional changes of shape of the MFs. This can be attributed to the effect of the increased width of the primary bump, as well as the secondary bumps that become increasingly more prominent, as $A_{\rm I}$ increases (see figure \ref{fig:bump_model}).
Crucially, MFs 
can distinguish bump models at the turnover scale 
$k^{\rm turn} = 0.5\,{\rm Mpc}^{-1}$ from the fiducial model, a 
regime where the models  are indistinguishable from the reionization history and globally averaged 21~cm regime~\cite{Naik:2025mba}. Importantly, this distinction 
arises because bump models modify the halo mass function across 
different mass scales (see figure~\ref{fig:hmf} in Appendix~\ref{app:hmf}), altering the distribution of collapsing 
structures even when averaged quantities such as 
$\langle x_{\rm HI} \rangle$ and $\langle \delta T_b \rangle$ remain 
unchanged. Thus, MFs capture morphological signatures inaccessible 
to global statistics, demonstrating their {power for constraining primordial features. In the next section, we examine  whether these inflationary signatures can be distinguished from variations in astrophysical parameters.

%%%%%%%%%%%%%%%%%%%%%%%%%%%%%%%%%%%%%%%%%%%%%%%%%%%%%%%%%%%%%%%%%%

\section{Distinguishability of primordial features from astrophysical parameters}
\label{sec:s5}

Our analysis of bump models using MFs has demonstrated 
their sensitivity to both the amplitude $A_\mathrm{I}$ 
and the location $k_\mathrm{peak}$ of primordial 
features. However, a key question arises for 
observational applications: can primordial parameters 
be constrained simultaneously with the astrophysical 
parameters governing the high-redshift universe? 
Specifically, do primordial features and astrophysical 
variations produce degenerate MF signatures? We address 
this question by comparing the morphological signatures 
of EoR parameter variations with those of bump models, 
identifying redshift regimes where the two can be 
distinguished.

%%%%%%%%%%%%%%%%%%%%%%%
\subsection{Effects of astrophysical parameters}
\begin{figure}[tbp]
    \centering
        (a) $\zeta$ variation\\
        \includegraphics[width=0.98\linewidth, height=4.5 cm]{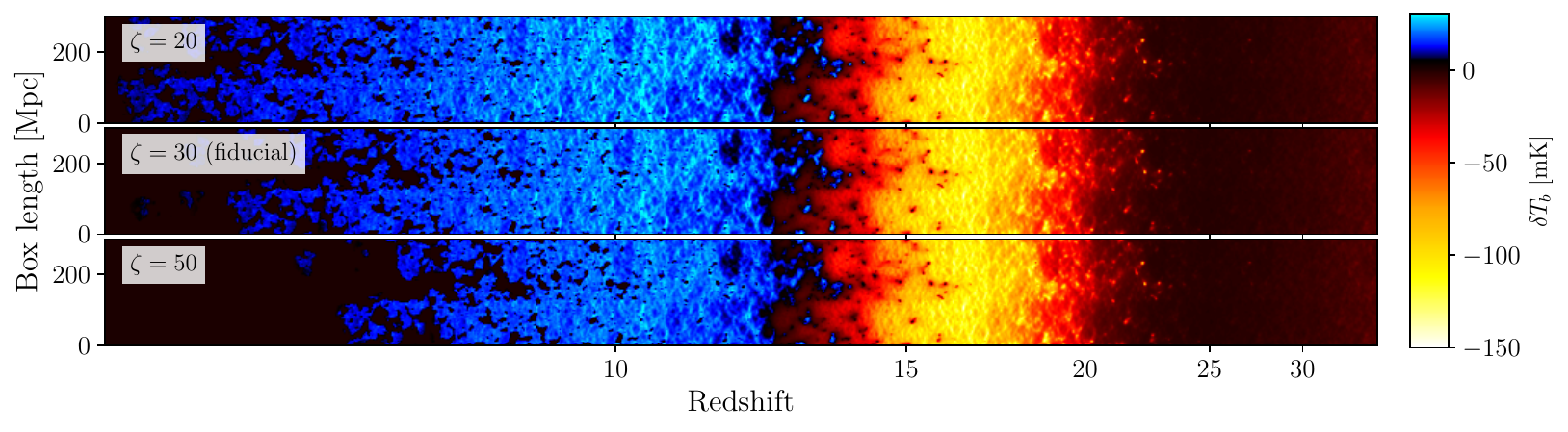} \\
        (b) $T_{\rm vir}$ variation \\
        \includegraphics[width=0.98\linewidth, height=4.5 cm]{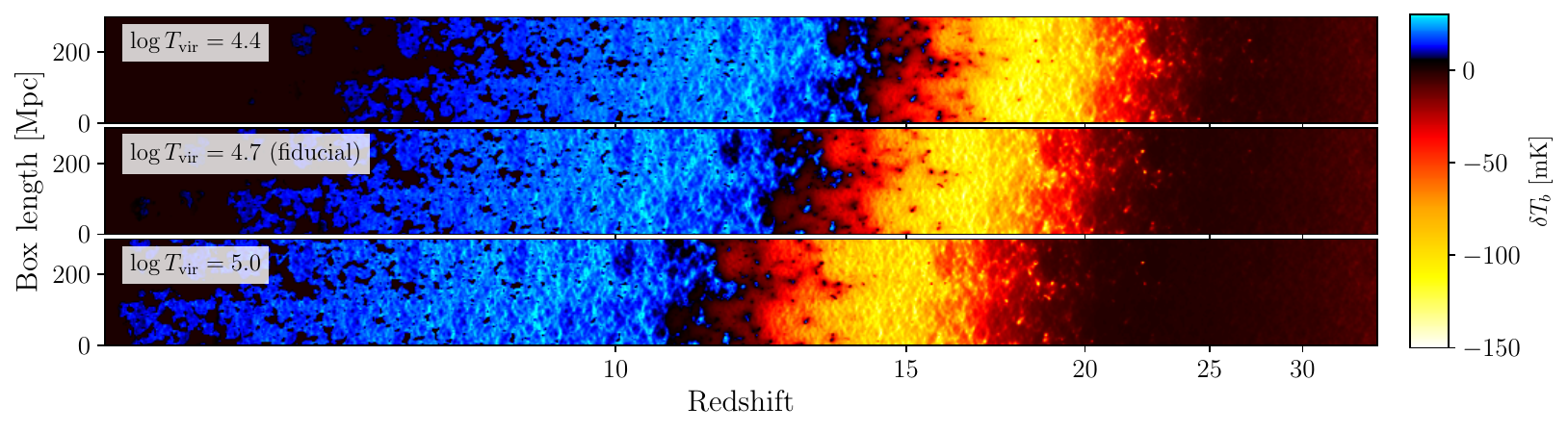} \\
        (c) $L_X$ variation \\
        \includegraphics[width=0.98\linewidth, height=4.5 cm]{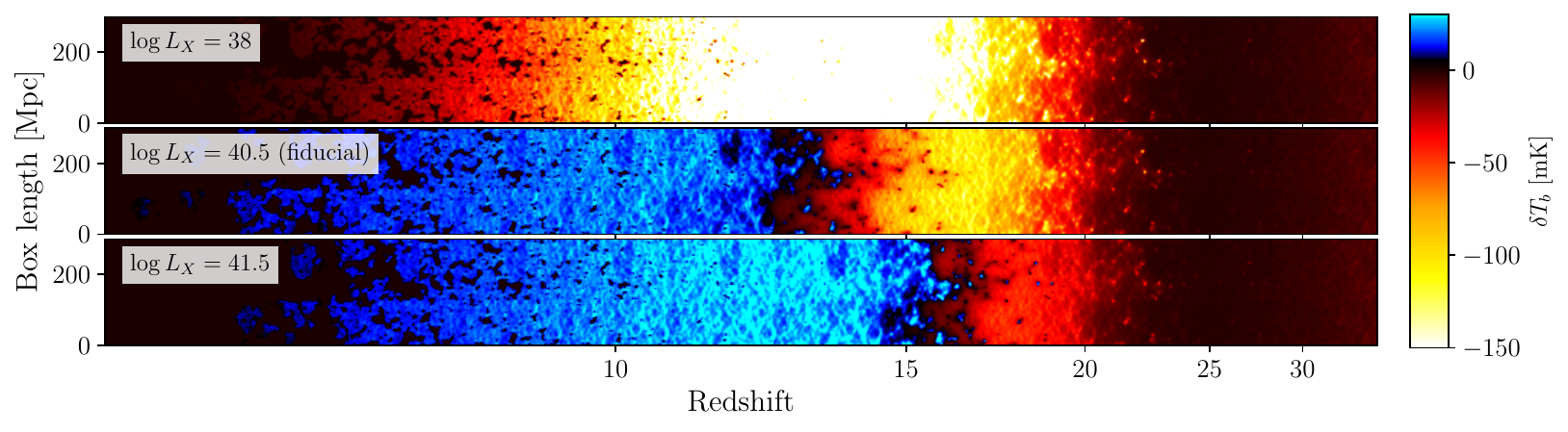}
    \caption{The simulated lightcones of the brightness temperature of the 21~cm signal for 
    variations in $\zeta$, $T_{\rm vir}$ and $L_X$.}
    \label{fig:lc_eor}
\end{figure}

In \texttt{21cmFAST}, the astrophysical processes during the dark 
ages and EoR are modelled by several parameters, with key parameters 
discussed in section~\ref{sec:s22}. We focus on three EoR parameters 
that significantly influence the ionisation history and consequently 
the evolution of the 21~cm signal: the ionising efficiency $\zeta$, 
the virial temperature threshold $T_{\rm vir}$, and the soft-band 
X-ray luminosity per unit SFR, $L_X$. The lightcones of the brightness 
temperature for variations in each of these parameters, in the range 
shown in table~\ref{tab:t1}, are presented 
in figure~\ref{fig:lc_eor}. We highlight the major effects below:
\begin{itemize}
    \item $\boldsymbol{\zeta}$: As shown in panel (a), $\zeta$ affects 
    the 21~cm brightness temperature at $z \lesssim 15$. Increasing 
    $\zeta$ enhances ionising photon production, accelerating 
    reionization and leading to earlier completion. Although bump 
    models with higher $k_{\rm peak}$ can also advance reionization, 
    unlike $\zeta$, the effects of $k_{\rm peak}$ are prominent at 
    $z > 15$ as well.
    \item $\boldsymbol{T_{\rm vir}}$: Since $T_{\rm vir}$ sets the 
    minimum halo mass for star formation, variations in $T_{\rm vir}$ 
    affect a broader redshift range than $\zeta$. Increasing 
    $T_{\rm vir}$ from $\mathcal{O}(10^4)$\,K to $\mathcal{O}(10^5)$\,K 
    shifts structure formation from being dominated by faint, low-mass 
    galaxies to bright, massive galaxies through modifications to the 
    HMF. Consequently, bump models may exhibit degeneracies with 
    $T_{\rm vir}$; indeed, analysis using the global 21~cm signal 
    revealed that $k_{\rm peak}$ produces degenerate effects with 
    $T_{\rm vir}$~\cite{Naik:2025mba}. We examine whether morphological 
    analysis can break this degeneracy below.
    \item $\boldsymbol{L_X}$: Variations in $L_X$ produce effects 
    that are highly distinguishable from the fiducial model. Since 
    X-ray luminosity heats the IGM, higher $L_X$ rapidly elevates 
    the kinetic temperature $T_K$ above $T_\gamma$, which couples 
    to $T_S$ and results in earlier emission signals. These effects 
    are particularly significant at $z \lesssim 20$.
\end{itemize}

%%%%%%%%%%%%%%%%%%%%%%%
\subsection{Quantifying morphological differences}

\begin{figure}[h!]
    \centering
    \includegraphics[width=.8\linewidth]{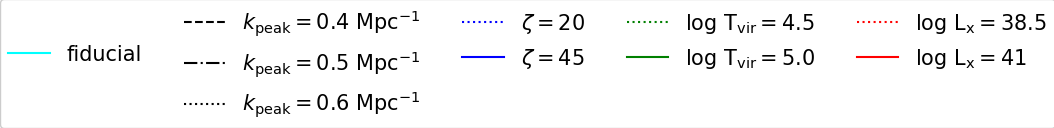}
    \includegraphics[width=1\linewidth]{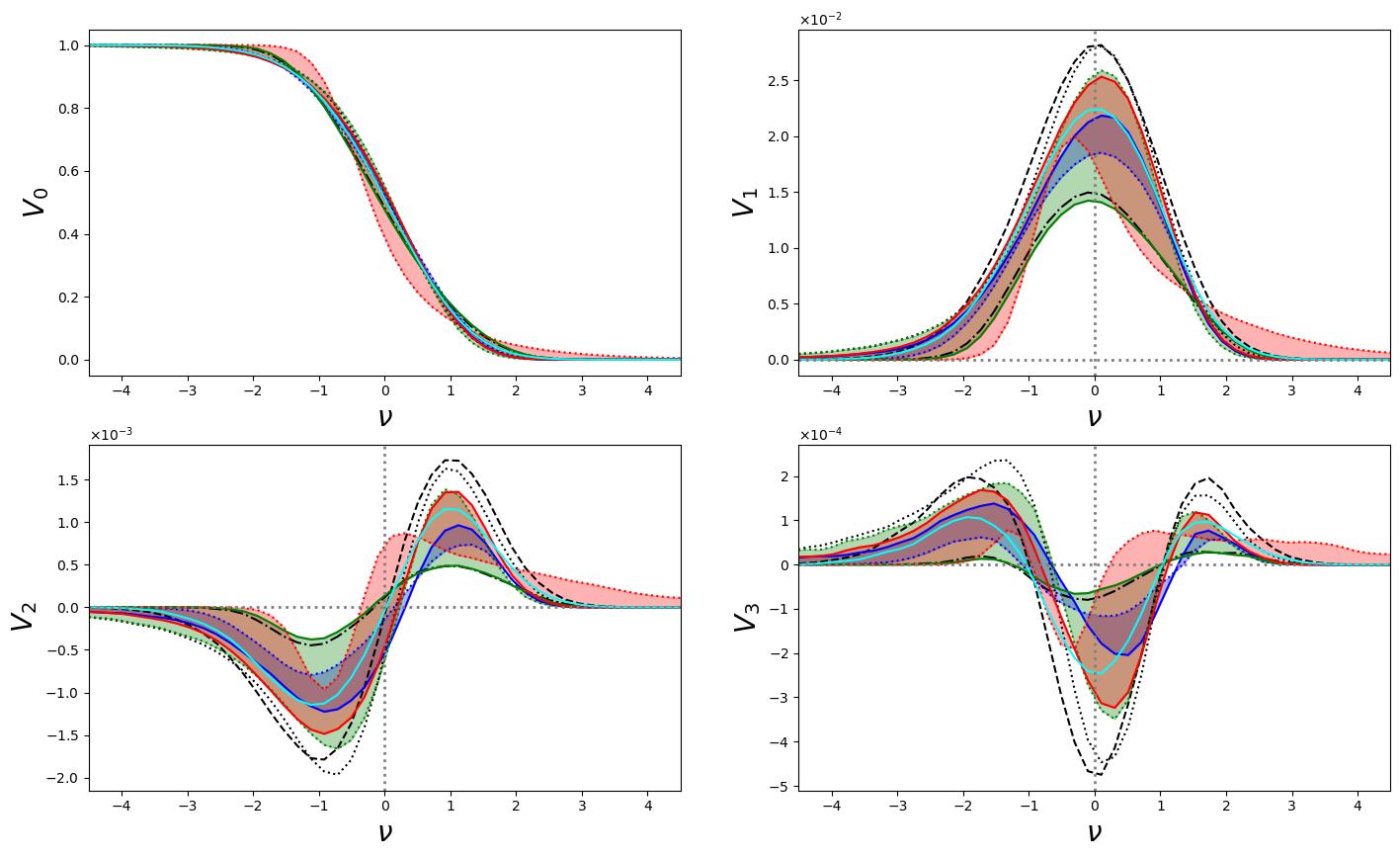}
    \caption{The MFs of the brightness temperature field for all models 
considered in this work, shown at $z = 11$. The solid cyan line 
represents the fiducial model, while dashed, dash-dotted, and dotted 
lines correspond to bump models with $k_{\rm peak} = 0.4$, $0.5$, 
and $0.6\,{\rm Mpc}^{-1}$, respectively. The shaded regions in blue, 
green, and red enclose the MFs spanning the full parameter ranges 
of $\zeta$, $T_{\rm vir}$, and $L_X$ listed in table~\ref{tab:t1}.}
    \label{fig:eor_char}
\end{figure}
%%%%%%%%%%%%%%%%%%%%%%%

We now jointly study the signatures of bump models and EoR scenarios to discern their distinct signatures, focusing only on the observable brightness temperature field. We vary the bump model 
parameter $k_{\rm peak}$ and EoR parameters as specified in table~\ref{tab:t1}. The bump amplitude is fixed to be $A_{\rm I}=10^{-8}$. All results shown in this section are for this value.  In figure~\ref{fig:eor_char}, we plot the MFs of the brightness temperature field for all models: fiducial, bump models, and EoR scenarios (with shaded regions spanning the range between minimum and maximum values of EoR parameters in table~\ref{tab:t1}) at redshift $z = 11$. %This redshift corresponds approximately to the midpoint of reionization for most models and thus includes contributions from all constituent fields in Eq.~\eqref{eq:brightness_temp}. 
%We select this value as a representative redshift for comparing the MF shapes across all EoR and bump models for the following reasons. 
We chose this redshift for visual representation simply because it roughly corresponds to the midpoint of reionization for most models and thus includes contributions from all constituent fields in 
eq.~\eqref{eq:brightness_temp}.
%At higher redshifts, reionization is unaffected by the $\zeta$ parameter, preventing meaningful comparison of MF variations for $\zeta$ models with others. At later redshifts, reionization is complete for some fast-reionizing models, causing their MF curves to flatten entirely. 
As can be seen visually,   %seen in Figure~\ref{fig:eor_char}, 
the shapes of $V_1$, $V_2$, and $V_3$ vary 
significantly across models, with visible differences 
in peak heights and curve asymmetry.

For $V_1$, the maximum near $\nu = 0$ has the highest 
amplitude for the $k_{\rm peak} = 0.6\,{\rm Mpc}^{-1}$ 
model, followed by other models in decreasing order, 
with the fiducial model exhibiting the lowest 
amplitude. The $V_1$ curves are not symmetric about 
$\nu = 0$; for example, the $\log{L_X} = 38.5$ curve 
is skewed toward negative thresholds, while the 
$\zeta = 45$ curve is skewed toward positive 
thresholds. This asymmetry can be quantified by 
comparing the areas under the curve on the positive 
and negative threshold sides. For $V_2$, the relative 
amplitudes of maxima and minima differ across models. 
The $k_{\rm peak} = 0.6\,{\rm Mpc}^{-1}$ curve 
exhibits the largest difference between its maximum 
and minimum, while the fiducial curve shows the 
smallest. Additionally, the $V_2$ curves are 
asymmetric about the %\sout{$y = 0$} 
$x$-axis line (horizontal black dotted line). For $V_3$, shown 
in the bottom right panel of figure~\ref{fig:eor_char}, 
the maxima at positive and negative threshold values 
have different amplitudes. For instance, in the 
$k_{\rm peak} = 0.6\,{\rm Mpc}^{-1}$ model, the 
maximum at negative thresholds has a higher amplitude 
than that at positive thresholds, whereas this 
behaviour is reversed for the 
$k_{\rm peak} = 0.4\,{\rm Mpc}^{-1}$ model. The $V_3$ curves are 
asymmetric about the %\sout{$y = 0$} 
$y$-axis line (vertical black dotted line).

These observations motivate the definition of derived 
measurable quantities by condensing the amplitude and shape information of 
$V_1$, $V_2$, and $V_3$, so as to extract the distinct signatures of different models. The amplitude quantifiers ${\cal A}_i$ are defined as:
\bea
 {\cal A}_1 &=&  V_1^{\rm max},\label{eq:A1} \\ 
 %%%
{\cal A}_2  &=& V_2^{\rm max} + |V_2^{\rm min}|, \label{eq:A2} \\
 %%%
{\cal A}_3 &=& \frac{V_3^{{\rm max},+} - V_3^{{\rm max},-}}{V_3^{{\rm max},+} + V_3^{{\rm max},-}},
\label{eq:A3} 
\eea
where $V_1^{\rm max}$  is the maximum of each $V_1$ curve,  $V_2^{\rm max/min}$ are the maximum and minimum respectively of each $V_2$ curve, and $V_3^{\rm max,+}$  ($V_3^{\rm max,-}$ ) is the maxima of each $V_3$ curve located at positive (negative) threshold.

To capture the shape information, we use integrals of the MF curves in suitable threshold ranges. To define the integration limits  we introduce the following special thresholds:  $\nu_{\rm min/max} = \mp 5$ which are the minimum and maximum thresholds considered here,  $\nu_c$ is the threshold where the $V_2$ curves cross zero (each curve crosses zero only once in the vicinity of $\nu=0$), and $\nu_{-}$ ($\nu_{+}$)  is the threshold value where the $V_3$ curves cross zero from positive to negative (negative to positive) values ($V_3$ curves cross zero at two points). These special thresholds vary from model to model. Then we define the shape quantifiers ${\cal E}_i$ as:
%%%
\bea
{\cal E}_1 &=& \int_{0}^{\nu_{\rm max}} V_1(\nu) \, d\nu - \int_{\nu_{\rm min}}^{0} V_1(\nu) \, d\nu,
\label{eq:E1} \\
%%%
{\cal E}_2 &=& \int_{\nu_c}^{\nu_{\rm max}} V_2(\nu)\, d\nu - \int_{\nu_{\rm min}}^{\nu_c} V_2(\nu) \, d\nu,
            \label{eq:E2}\\
%%%
{\cal E}_3 &=& \frac{ \bigg(\int_{\nu_{+}}^{\nu_{\rm max}} V_3(\nu) \, d\nu - \int_{\nu_{\rm min}}^{\nu_-} V_3(\nu) \, d\nu\bigg)}{ \bigg(\int_{\nu_+}^{\nu_{\rm max}} V_3(\nu) \, d\nu + \int_{\nu_{\rm min}}^{\nu_-} V_3(\nu) \, d\nu\bigg)} .  
\label{eq:E3}
\eea
%%%%%%%%
These six parameters ${\cal A}_i,\,{\cal E}_i$ probe different aspects of the field morphology: ${\cal A}_i$  measure information about amplitudes of the MFs and which are related to structural scales, while ${\cal E}_i$ measures information associated with asymmetries and shapes of the MFs. Note that we do not include information from $V_0$ since it is known to have large cross-correlations across different thresholds relative to other MFs,  rendering it difficult to construct its likelihood function and covariance matrix (see ref.~\cite{Bashir:2025} for details).

%\red{Describe how error bars are computed.} 
For the estimation of statistical uncertainties, we generate 30 simulations with different random seeds for each model. MFs are calculated from all the simulations, and then the means and standard deviations of all the derived quantities defined in eqs.~\ref{eq:A1} to \ref{eq:E3} are computed. This is plotted in the upper panels of figure \ref{fig:maxima_params} and \ref{fig:area_params}.
To quantify the distinguishability of the different models, for each derived statistic at each redshift, we calculate the distance of the mean value for each model from the mean of the fiducial model in units of the standard deviation of the field.  These distances are shown in the lower panels of figure \ref{fig:maxima_params} and \ref{fig:area_params}. The shaded yellow bands mark the $5\sigma$ deviations from the fiducial model. We use this as a benchmark when we describe distinguishability of different models below.
%The models outside this band are distinguishable from the fiducial model.  

 %%%%%%%%%%%%%%
\begin{figure}[t]
    \centering
    \includegraphics[width=0.7\linewidth]{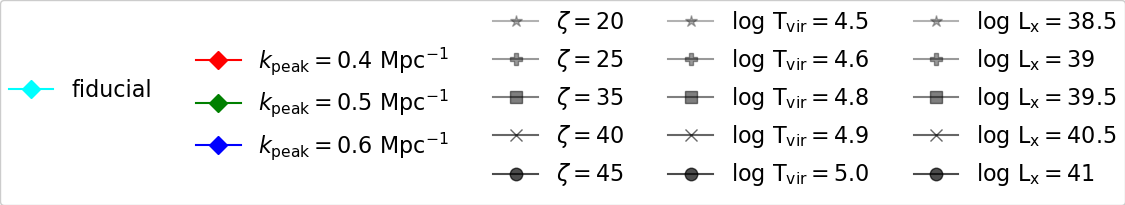}
    \includegraphics[width=0.95\linewidth]{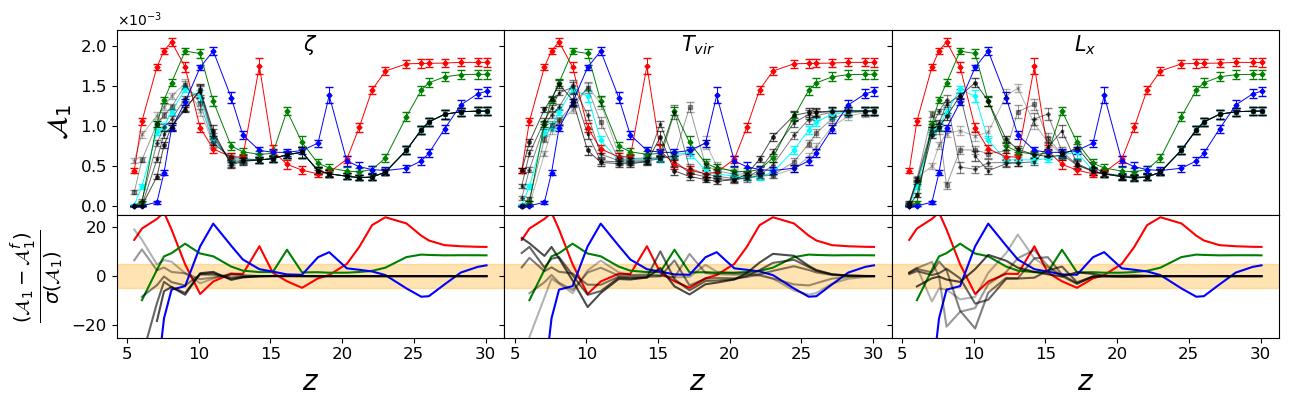}
    \includegraphics[width=0.95\linewidth]{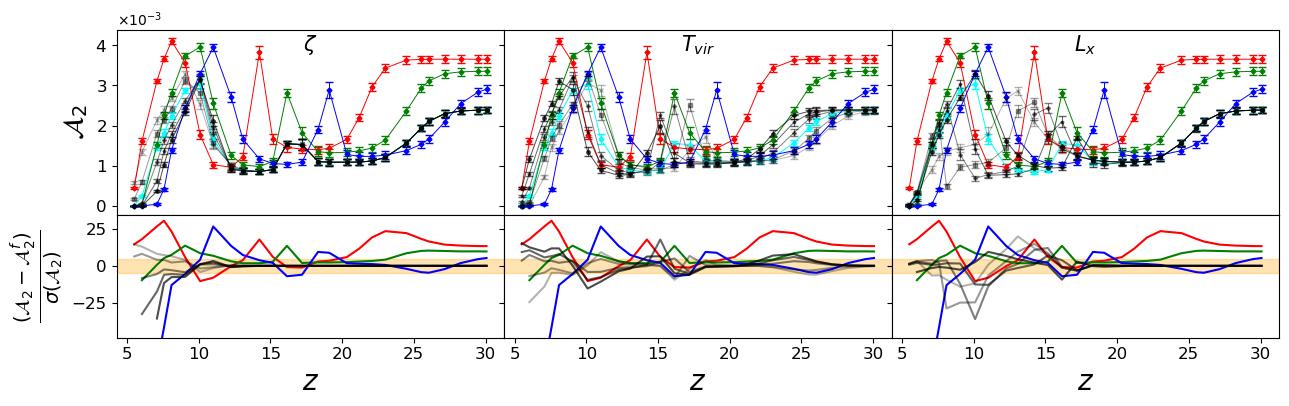}
    \includegraphics[width=0.95\linewidth]{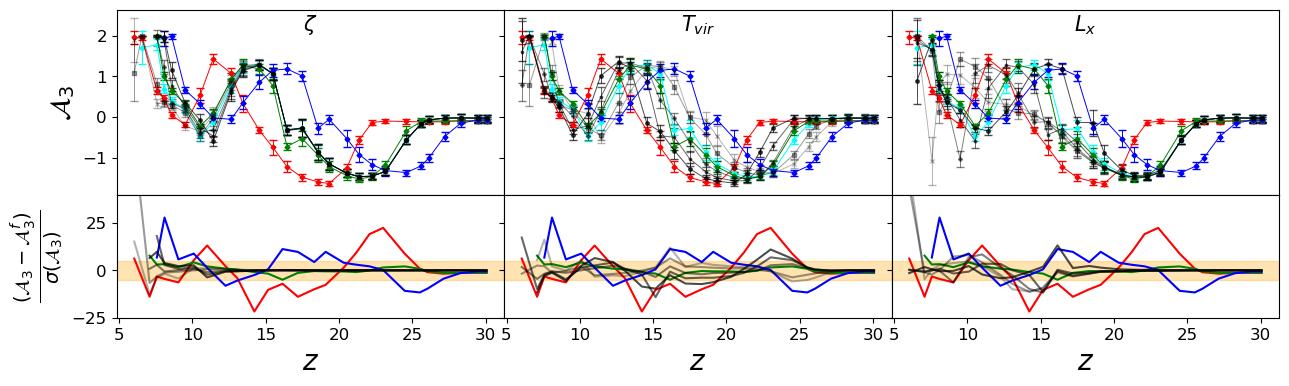}
    \caption{Redshift evolution of the amplitude parameters 
    $\mathcal{A}_1$, $\mathcal{A}_2$, and $\mathcal{A}_3$ 
    (defined in eqs.~\ref{eq:A1}--\ref{eq:A3}) for bump 
    models and EoR parameter variations. The columns correspond 
    to variations in $\zeta$ (left), $T_{\rm vir}$ (centre), and 
    $L_X$ (right). The upper panels in each row show the parameter 
    values with $1\sigma$ error bars estimated from 30 realisations. 
    The lower panels show the deviation of each model from the 
    fiducial in units of the standard deviation. The colours and 
    markers for bump models and EoR models are as indicated in 
    the legend.}
    \label{fig:maxima_params}
\end{figure}
%%%%%%%%%%%%%%%%%
\begin{figure}[t]
    \centering
    \includegraphics[width=0.7\linewidth]{figures/legend6.png}
    \includegraphics[width=0.95\linewidth]{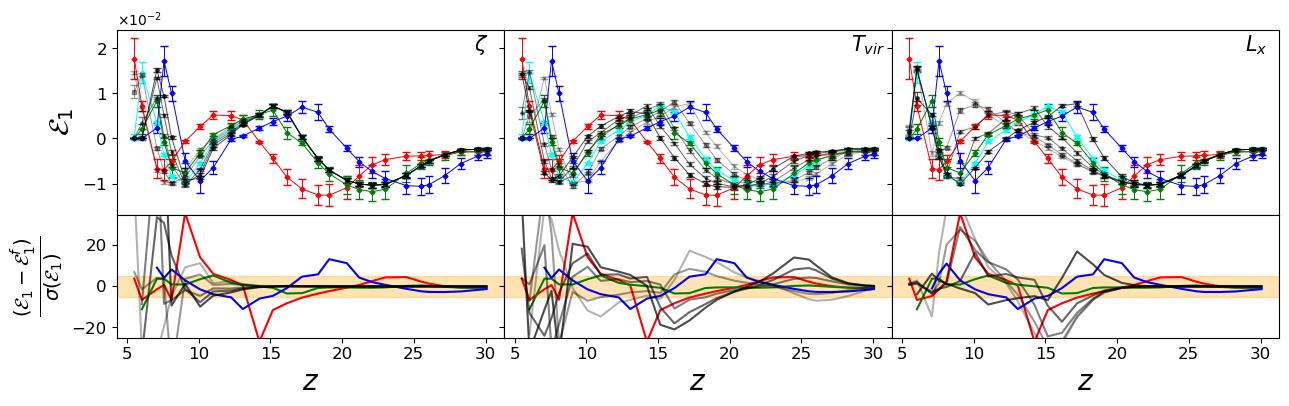}
    \includegraphics[width=0.95\linewidth]{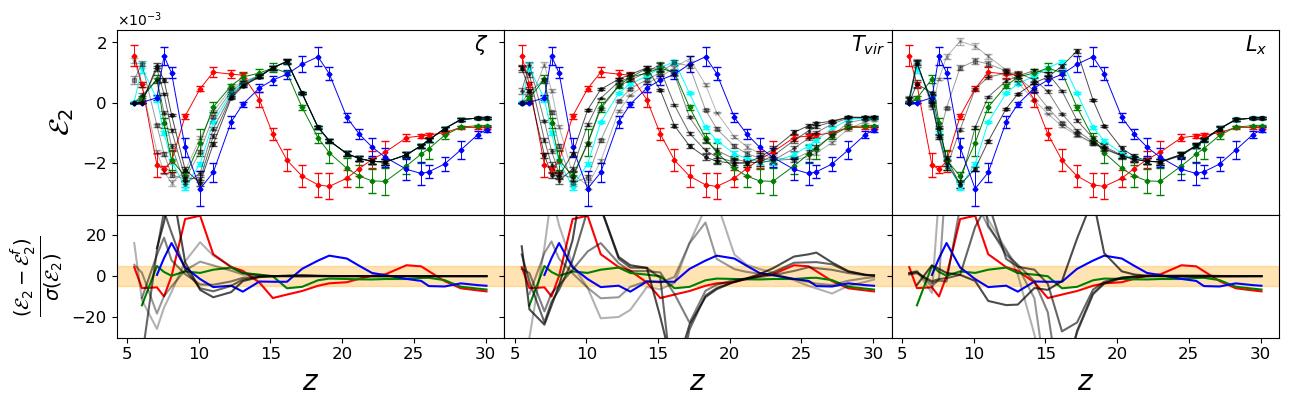}
    \includegraphics[width=0.95\linewidth]{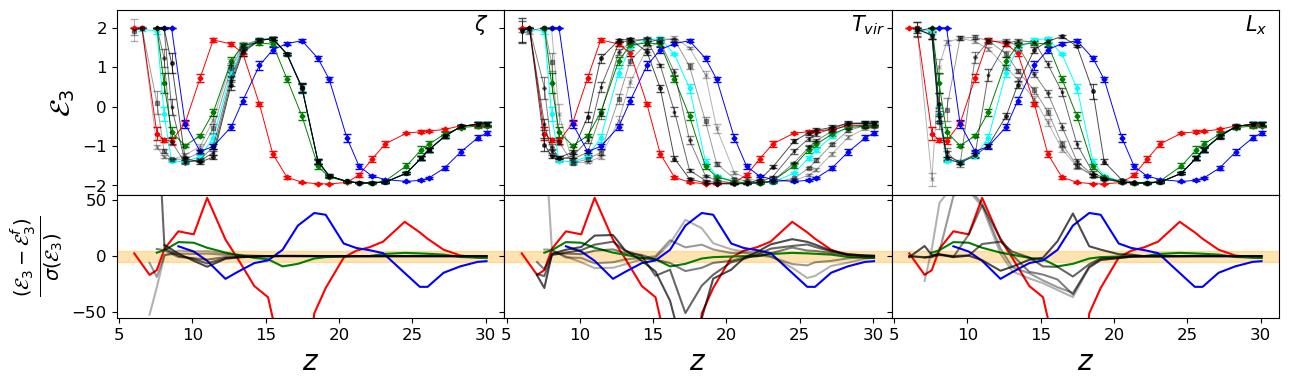}
    \caption{Same as figure~\ref{fig:maxima_params}, but for the 
    shape parameters $\mathcal{E}_1$, $\mathcal{E}_2$, and 
    $\mathcal{E}_3$ (defined in eqs.~\ref{eq:E1}--\ref{eq:E3}).}
    \label{fig:area_params}
\end{figure}

We now discuss the salient features of figures \ref{fig:maxima_params} and \ref{fig:area_params} in three redshift ranges - high, intermediate and low. The ranges are chosen based on the dominant contributions to $\langle \delta T_b\rangle$ from different terms in eq.~\eqref{eq:brightness_temp}, as explained in section \ref{sec:s22} (see figure \ref{fig:f2}). % For brevity we drop the ${\rm Mpc}^{-1}$ unit when we refer to $k_{\rm peak}$. }

\paragraph{High redshifts ($z > 20$):} At these high redshifts, the 21 cm signal is dominated by density field, as ionising sources are sparse. Consequently, most bump models are distinguishable from both fiducial and EoR models.  
Among the amplitude quantifiers, $\mathcal{A}_1$ and $\mathcal{A}_2$ exhibit similar behaviour:  the $k_{\rm peak} = 0.4\,{\rm Mpc}^{-1}$  
model becomes distinguishable at higher than $5\sigma$ significance beyond $z = 20$, the $k_{\rm peak} = 0.5\,{\rm Mpc}^{-1}$ 
model beyond $z = 22$, and the $k_{\rm peak} = 0.6\,{\rm Mpc}^{-1}$ model 
beyond $z = 27$. For $\mathcal{A}_3$, the $k_{\rm peak} = 0.4\,{\rm Mpc}^{-1}$ 
model is distinguishable in the range $21 < z < 26$, while $k_{\rm peak} = 0.6\,{\rm Mpc}^{-1}$ 
is distinguishable for $23 < z < 28$.
Among the shape quantifiers,  $\mathcal{E}_2$ shows that both the $k_{\rm peak} = 0.5$ and $0.6\,{\rm Mpc}^{-1}$
models become distinguishable beyond $z = 27$. For $\mathcal{E}_3$, the $k_{\rm peak} = 0.4\,{\rm Mpc}^{-1}$ 
model is distinguishable in the range $22 < z < 28$, and $k_{\rm peak} = 0.6\,{\rm Mpc}^{-1}$ 
beyond $z = 23$. Of the EoR parameters, $T_{\rm vir}$ shows degeneracy with $k_{\rm peak}$ in this redshift range. This high-redshift cosmic dawn regime offers the cleanest separation between primordial and astrophysical signatures, making it %\sout{optimal} 
very well suited %\red{$<--$ I cut optimal since it is a technical work and needs quantification} 
for constraining bump model parameters.

%\paragraph{High redshifts ($z > 20$):} At these high redshifts, the 21 cm signal is dominated by density fluctuations, as ionising sources are sparse. Consequently, most bump models are distinguishable from both fiducial and EoR models.  
    %
%Among the amplitude quantifiers, $\mathcal{A}_1$ and $\mathcal{A}_2$ exhibit similar behaviour:  \mg{the B${}_-$ model} becomes distinguishable \mg{at higher than $5\sigma$ significance} beyond $z = 20$, \mg{the B${}_0$ model} beyond $z = 22$, and \mg{the B${}_+$ model}  beyond $z = 27$. For $\mathcal{A}_3$, the $k_{\rm peak} = 0.4\,{\rm Mpc}^{-1}$ model is distinguishable in the range $21 < z < 26$, while $k_{\rm peak} = 0.6\,{\rm Mpc}^{-1}$ is distinguishable for $23 < z < 28$.
    %
%Among the shape quantifiers,  $\mathcal{E}_2$ shows that both the $k_{\rm peak} = 0.5$ and $0.6\,{\rm Mpc}^{-1}$ models become distinguishable beyond $z = 27$. For $\mathcal{E}_3$, the $k_{\rm peak} = 0.4\,{\rm Mpc}^{-1}$ model is distinguishable in the range $22 < z < 28$, and $k_{\rm peak} = 0.6\,{\rm Mpc}^{-1}$ beyond $z = 23$. This high-redshift \mg{cosmic dawn} regime offers the cleanest separation between primordial and astrophysical signatures, making it \sout{optimal} \mg{very well suited} for constraining bump model parameters.

    %
\paragraph{Intermediate redshifts ($10 < z < 20$):}
This regime spans the active phase of reionization, where the 21\,cm signal receives contributions from all constituent fields. The interplay between primordial features and astrophysical processes  makes the distinguishability of bump models more nuanced and strongly redshift-dependent.    
For the amplitude quantifiers $\mathcal{A}_1$ and $\mathcal{A}_2$, the turnover-scale model ($k_{\rm peak} = 0.5\,{\rm Mpc}^{-1}$) 
is distinguishable only within a narrow window around $z \sim 16$, while the $k_{\rm peak} = 0.6\,{\rm Mpc}^{-1}$ 
model shows distinguishability in two separate windows: $10 < z < 12$ and $18 < z < 20$. For $\mathcal{A}_3$, the $k_{\rm peak} = 0.4\,{\rm Mpc}^{-1}$ 
model is distinguishable over a broader range $13 < z < 19$, whereas $k_{\rm peak} = 0.6\,{\rm Mpc}^{-1}$ 
is distinguishable for $16 < z < 20$.
For the shape quantifiers, $\mathcal{E}_1$ shows that $k_{\rm peak} = 0.4\,{\rm Mpc}^{-1}$ 
is distinguishable for $13 < z < 17$ and $k_{\rm peak} = 0.6\,{\rm Mpc}^{-1}$
for $16 < z < 20$. In $\mathcal{E}_2$, the $k_{\rm peak} = 0.4\,{\rm Mpc}^{-1}$ 
model shows the broadest window of distinguishability ($14 < z < 20$), whereas $k_{\rm peak} = 0.5\,{\rm Mpc}^{-1}$ 
is distinguishable only near $z \sim 17$ and $k_{\rm peak} = 0.6\,{\rm Mpc}^{-1}$ 
for $18 < z < 20$. For $\mathcal{E}_3$, the $k_{\rm peak} = 0.4\,{\rm Mpc}^{-1}$ 
model exhibits two windows of distinguishability ($10 < z < 13$ and $14 < z < 17$), the turnover-scale model is distinguishable only at $z \sim 17$, and the $k_{\rm peak} = 0.6\,{\rm Mpc}^{-1}$
model for $17 < z < 20$.
    % In amplitude measures -  ${\cal A}_1$ and ${\cal A}_2$, $k_{\rm peak} = 0.5$ is distinct in the narrow window around $z=16$ and $k_{\rm peak} = 0.6$ in $z=10 \ \rm to \ 12$ and $z=18 \ \rm to \ 20$. For ${\cal A}_3$, $k_{\rm peak} = 0.4$ is distinguishable in $13<z<19$, $k_{\rm peak} = 0.6$ in $16<z<20$. In area measures - for ${\cal E}_1$, $k_{\rm peak} = 0.4$ is distinct in the window $13<z<17$ and $k_{\rm peak} = 0.6$ is distinct in the window $16<z<20$. for ${\cal E}_2$, $k_{\rm peak} = 0.4$ is distinct for $14<z<20$, $k_{\rm peak} = 0.5$ around $z = 17$ and $k_{\rm peak} = 0.6$ in $18<z<20$. In ${\cal E}_3$, $k_{\rm peak} = 0.4$ is distinct in the range $10<z<13$ and $14<z<17$. $k_{\rm peak} = 0.5$ is distinct at redshift $z=17$ and $k_{\rm peak} = 0.6$ is distinct in the window $18<z<20$.
    %
\paragraph{Low redshifts ($z < 10$):} %At these \sout{late stages} 
In this redshift range, the 21\,cm signal morphology is dominated by ionised regions surrounding luminous 
sources. Variations in both $\zeta$ and $L_X$ 
significantly affect the signal, leading to 
substantial overlap between EoR and bump model 
signatures. Consequently, bump models are 
distinguishable from EoR scenarios only in limited 
cases.
For the amplitude quantifiers $\mathcal{A}_1$ and 
$\mathcal{A}_2$, only the 
$k_{\rm peak} = 0.4\,{\rm Mpc}^{-1}$ model remains 
distinguishable, and only below $z \simeq 8$. For 
$\mathcal{A}_3$ and the shape quantifier 
$\mathcal{E}_2$, the 
$k_{\rm peak} = 0.6\,{\rm Mpc}^{-1}$ model shows 
marginal distinguishability around $z \sim 8$.

%\blue{Bump models are most robustly distinguished from EoR variations at $z > 20$, where density fluctuations dominate and astrophysical processes have minimal impact. At intermediate redshifts ($10 < z < 20$), distinguishability is parameter-dependent, with the turnover-scale model ($k_{\rm peak} = 0.5\,{\rm Mpc}^{-1}$) identifiable only in narrow redshift windows. At $z < 10$, overlap with EoR signatures limits the constraining power of MFs for primordial features. Notably, combining multiple quantifiers across redshifts can enhance the overall constraining  power, as different $k_{\rm peak}$ values show complementary windows of distinguishability.}

To summarize, bump models are robustly distinguished from EoR models at $z > 20$, where density fluctuations dominate and astrophysical processes have minimal impact. At intermediate redshifts ($10 < z < 20$), model sensitivities  vary with the statistics used, with the turnover-scale model ($k_{\rm peak} = 0.5\,{\rm Mpc}^{-1}$) identifiable in specific redshift windows. At relatively low redshift ($z < 10$), we find degeneracy of bump and EoR parameters. However, there exists redshift windows where they can be distinguished for different statistics. Notably, combining multiple statistics across redshifts can enhance the overall constraining  power, as different $k_{\rm peak}$ values show complementary windows of distinguishability.

%\mg{As previously mentioned, the above results are for the bump amplitude value $A_{\rm I}=10^{-8}$. The sensitivity of the MFs to model parameters is expected to decrease with decrease of  $A_{\rm I}$.}

%\red{Points to add:\\
%1.  Comment on what will happen to distinguishability for $A_{\rm I}$ value.\\
%2. Which statistic amongst ${\cal A}_i$ and ${\cal E}_i$ exhibit most distinguishability. In general ${\cal E}_i$  is better than ${\cal A}_i$, and among them ${\cal E}_3$ shows strongest sensitivity.\\
%3. MFs can constraining the turnover model because of non-Gaussianity of the brightness temperature. }

%%%%%%%%%%%%%%%%%%%%%%%%%%%%%%%%%%%%%%%%%%%%%%%%%%%%%%%%%%%%%%%%%%%%%%%%
\section{Summary and Discussion}
\label{sec:s6}
% Point 1: What we have done, what is unique and important
In this work, we have explored the morphological signatures of bump-like primordial features arising from particle production during inflation on the 21~cm signal from cosmic dawn and the Epoch of Reionisation. The bump model is parameterised by the amplitude $A_{\rm I}$ and the scale corresponding to the peak of the feature, $k_{\rm peak}$. We modified the initial power spectrum in the semi-numerical simulation code \texttt{21cmFAST} to incorporate bump-like features and compared the resulting fields with a fiducial model that adopts the nearly scale-invariant power spectrum.

Our analysis proceeded in two parts. First, we demonstrated that  Minkowski functionals of the 21~cm signal are sensitive to primordial bump-like features. We compared six bump models with three $k_{\rm peak}$ values -- $0.4$, $0.5$, and $0.6\,{\rm Mpc}^{-1}$ and two $A_{\rm I}$ values  -- $10^{-9}$ and $10^{-8}$ with the fiducial case.  We systematically examined how bump features modify the morphology of the constituent cosmological fields -- density field $\delta_b$, spin temperature $T_S$, and neutral hydrogen fraction $x_{\rm HI}$ -- that collectively determine the morphology of the 21~cm brightness temperature $\delta T_b$. We found that bump-like features enhance overdensities at scales corresponding to $k_{\rm peak}$, modifying the characteristic  length scale of density structures and thereby increasing MF amplitudes. These modifications propagate through the halo mass function to affect the timing of structure formation, heating, and reionisation, leaving distinct imprints on the MFs of $T_S$, $x_{\rm HI}$, and $\delta T_b$ across different redshift regimes. Importantly, MFs can distinguish bump models 
at the turnover scale $k^{\rm turn} = 0.5\,{\rm Mpc}^{-1}$ from 
the fiducial model -- a scale where the global 21~cm signal and 
reionisation history are indistinguishable~\cite{Naik:2025mba}. 
% This demonstrates that MFs capture morphological information 
% inaccessible to globally averaged statistics. 

In the second part, we investigated potential degeneracies between primordial signatures and astrophysical uncertainties by considering  different EoR scenarios based on three key parameters: $\zeta$, $T_{\rm vir}$, and $L_X$. By defining measurable quantities derived from MFs that quantify amplitude ($\mathcal{A}_1$, $\mathcal{A}_2$, $\mathcal{A}_3$) and shape ($\mathcal{E}_1$, $\mathcal{E}_2$, $\mathcal{E}_3$) %\sout{parameters  derived from the MFs} 
information (eqs.~\eqref{eq:A1} -- \eqref{eq:E3}), we identified redshift ranges where bump models can be distinguished from EoR models. 
We found that bump models are most robustly 
distinguished at $z > 20$ across all statistics, where density fluctuations dominate 
and astrophysical processes have minimal impact. At intermediate 
redshifts ($10 < z < 20$) and lower redshifts ($z < 10$) we found redshift windows, depending on the specific statistic and $k_{\rm peak}$ value, where the models can be distinguished. 
%and distinguishability depends on the specific statistic and $k_{\rm peak}$ value, while at $z < 10$, overlap with EoR signatures limits the constraining power. 
Such derived statistics can serve as statistics for model selection by comparing with observed data using a full Markov Chain Monte Carlo (MCMC).

% Point 2: Importance in context of experiments
Our results demonstrate 
that MFs carry characteristic signatures of primordial bump-like 
features that are distinct from both the fiducial model and EoR 
parameter variations. While the 21~cm power spectrum is also sensitive to bump-like features~\cite{Naik:2022wej}, MFs provide complementary information by probing the non-Gaussian morphological properties of the signal; a detailed analysis of bump model signatures in the 21~cm power spectrum will be presented in forthcoming work (Naik et al., in preparation). Importantly, 21~cm observations across multiple redshifts are essential for robustly identifying these signatures, as different redshift regimes provide complementary information. The upcoming SKA will map the 21~cm brightness temperature in three dimensions over the redshift range $6 < z < 30$, providing precisely the tomographic data required for morphological analysis. 

% Point 3: Shortcomings and future directions
This work represents a first step towards using MFs to probe 
primordial features in the 21~cm signal and extricate the physical implications. While we have accounted for sample variance by averaging over multiple realisations, our 
analysis uses idealised simulations without observational effects 
such as thermal noise and foreground contamination; incorporating 
these will be essential for assessing detectability with real data. 
Although we focused on bump-like features from particle production 
during inflation, this is the first study to explore the imprints 
of primordial features on the morphology of the 21~cm signal. The 
framework developed here is general and can be readily applied to 
other inflationary models predicting scale-dependent features.
Such studies are important, 
as detection of primordial features would provide insights 
into the physics of the inflationary epoch beyond what is accessible 
from CMB observations alone.

In future works we plan to extend this analysis in several 
directions. We will incorporate %\sout{more}
realistic SKA mock data 
to assess the detectability of bump model signatures under 
observational conditions. We will also perform a full Bayesian 
inference analysis using MCMC methods 
to quantify the constraining power of MFs for the bump model 
parameters $A_{\rm I}$ and $k_{\rm peak}$, jointly exploring 
the EoR parameter space to provide a more complete picture of 
potential degeneracies. Additionally, comparing with other 
statistics such as the power spectrum and bispectrum will further 
enhance the ability to disentangle primordial and astrophysical 
signatures. 
These developments will establish the full potential of morphological 
statistics for probing inflationary physics with upcoming 
21~cm surveys.

\section*{Acknowledgements}
\label{sec:ack}
%%%%%%%%%%%%%%%%%%%%%%%%%%
The computational work in this paper was carried out using the NOVA cluster at the Indian Institute of Astrophysics, Bengaluru. The authors thank Kazuyuki Furuuchi for a careful reading of the manuscript and helpful comments.
SSN acknowledges support from the SERB-NPDF (File No. PDF/2023/001469), Anusandhan National Research Foundation, Government of India.

%%%%%%%%%%%%%%%%%%%%%%%%%%%%%%%%%%%%%%%%%%%%%%%%%%%%%%%%%%%%%%%%%%
\appendix
%%%%%%%%%%%%%%%%%%%%%%%%%%%%%%%%%%%%%%%%%%%%%%%%%%%%%%%%%%%%%%%%%%
\section{Effect of primordial bump-like features on the halo mass function} 
\label{app:hmf}
%%%%%%%%%%%%%%%%%%%%%%%%%%%%%%%%%%%%%%%%%%%%%%%%%%%%%%%%%%%%%%%%%%
In section~\ref{sec:s41}, we discussed how primordial bump-like features affect 
21~cm signals through two mechanisms: (i) by modifying the density field 
via the matter power spectrum, and (ii) by influencing the neutral 
hydrogen and spin temperature fields through changes in the collapsed fraction of halos, 
which depends on the halo mass function (HMF). 
In this section, we provide a concise summary of the second mechanism. 
For a comprehensive analysis, we refer the reader to \cite{Naik:2025mba}. 

The halo mass function, $\frac{dn}{d\ln M}\, [{\rm Mpc}^{-3}]$, 
quantifies the comoving number density of halos 
in the mass range $M$ to $(M+dM)$, and 
can be written as \cite{Press_Schechter1974ApJ...187..425P} 
\begin{equation}
    \frac{dn}{d \ln M} 
    =
    \frac{\rho_m}{M}
    \frac{-d(\ln \sigma)}{dM}
    \nu 
    f(\nu)
    \,,
    \label{eq:dndM}
\end{equation}
where 
$\rho_m$ represents the average matter density at $z=0$ 
and 
$\nu$ is the critical overdensity threshold. 
% and 
% $\nu \equiv  \frac{\delta_c}{D(z) \sigma(M)}$, 
% where $\delta_c$ is the critical overdensity, 
% $D(z)$ is the growth factor. 
In \texttt{21cmFAST}, the Sheth-Tormen formula \cite{Sheth:2001dp} 
for the HMF is used to 
calculate the number density of halos via $\nu f(\nu)$. 
The variance of the initial density fluctuation field, $\sigma(M)$, is obtained
by linearly extrapolating to the present epoch and smoothing with a filter $W(kR)$ of scale $R$, i.e., 
\begin{equation}
    \sigma^2(R) =
    \frac{1}{2\pi^2}
    \int_0^\infty 
    dk \,
    k^2 P_m(k)
    W^2(kR)
    \,,
    \label{eq:smoothed_sigma}
\end{equation}
where $P_m(k)$ is the matter power spectrum. 
In \texttt{21cmFAST}, a real-space top-hat filter is used by default, 
whose Fourier transform is given by 
\begin{equation}
    W(kR) = 3 \left[\frac{\sin(kR) - (kR) \cos(kR)}{(kR)^3}\right]\,,
    \label{eq:filter}
\end{equation}
where the radius of the top-hat filter, $R$, and the enclosed mass $M$
are related by
\begin{equation}
    R (M) \equiv 
    \left[\left(\frac{3}{4\pi}\right)
    \left(\frac{M}{\rho_m}\right)\right]^{1/3}\,.
    \label{eq:tophat}
\end{equation}

\begin{figure}[tbp]
    \centering
    \includegraphics[width=\linewidth]{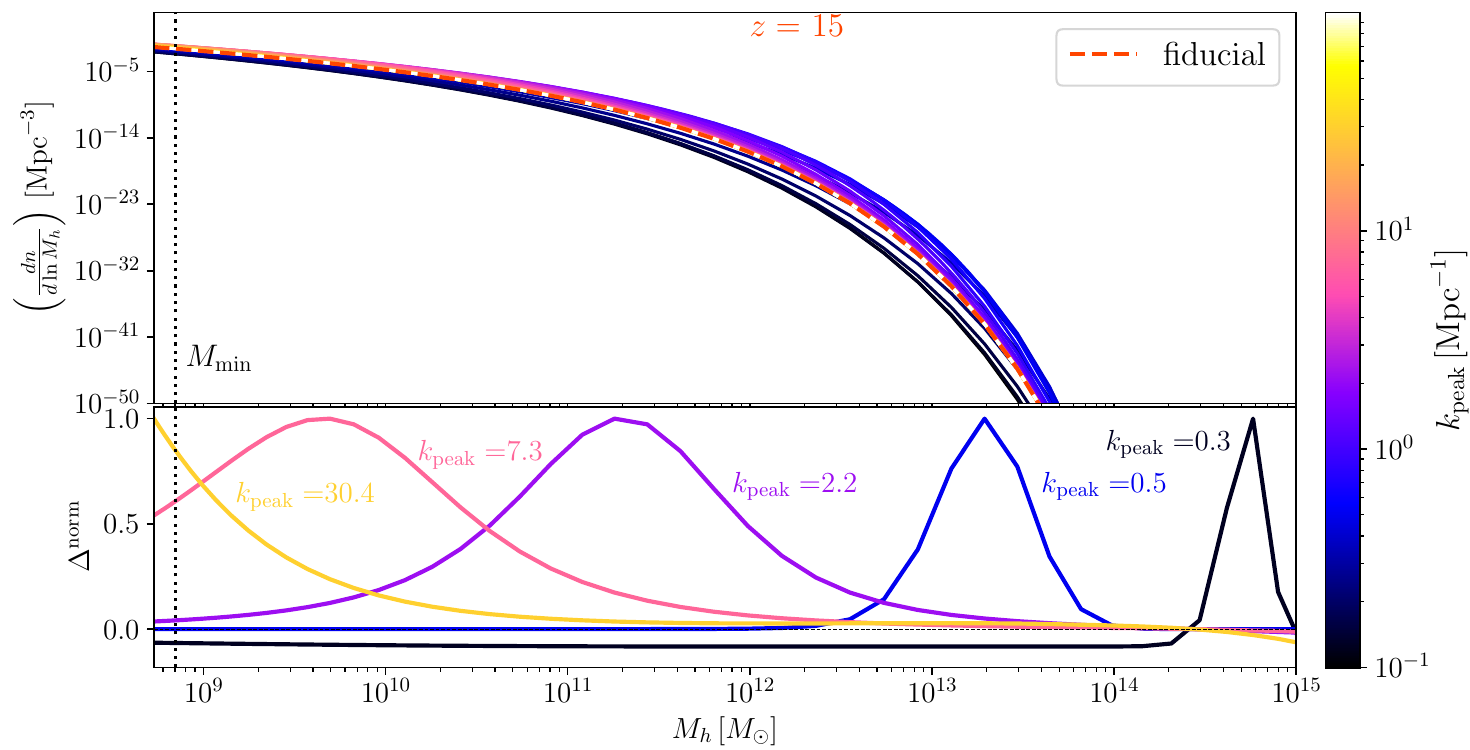} 
    \caption{[Top] The halo mass function at $z = 15$, simulated using \texttt{21cmFAST}
    for various bump models, with $k_{\rm peak}$ values shown in the color bar and 
    the fiducial model indicated by the dashed line.
    [Bottom] The normalized relative difference of the HMF for the bump models 
    with respect to the fiducial model, $\Delta^{\rm norm}$, for 
    several values of $k_{\rm peak}$. 
    The vertical dashed line corresponds to the mass scale $M_{\rm min}$.}
    \label{fig:hmf}
\end{figure}
From the above equations, it is evident that modifications 
to the power spectrum enter the halo mass function via the term 
$\sigma^2(M)$. 
The resulting halo mass functions for the bump models as simulated using \texttt{21cmFAST}
are plotted in figure~\ref{fig:hmf}.  
On the bottom panel, we also show $\Delta^{\rm norm}$ -- the relative difference between the bump and fiducial models, 
normalized by the maximum, for several values of $k_{\rm peak}$. 
The plots indicate that bumps with lower values of $k_{\rm peak}$
increase the number density of high-mass halos and 
decrease the number density of low-mass halos. 
With increasing values of $k_{\rm peak}$, 
the enhancement in the HMF shifts to lower mass scales. 
This anti-correlation between $k_{\rm peak}$ and $M_{\rm peak}$ (the mass corresponding 
to the peak in the HMF curve) arises from the fundamental relationship between wavenumbers and mass:  
a primordial bump-like feature with low $k_{\rm peak}$
affects large-scale fluctuations, and larger length scales correspond to larger mass scales 
(see eq.~\eqref{eq:tophat}). 

The astrophysical implications of these HMF modifications become clear when considering 
the vertical dashed line in the figure, which corresponds to the mass scale $M_{\rm min}$
that sets the threshold above which halo collapse becomes important for reionization. 
Since low-mass halos in the vicinity of $M_{\rm min}$ dominate the collapse fraction
and drive early structure formation, variations in their number density 
directly impact the timing of reionization. 
Consequently, a bump model with low $k_{\rm peak}$ (e.g., 0.3) 
decreases the number density of these critical low-mass halos, resulting in 
a lower collapsed fraction relative to the fiducial model and 
delayed structure formation and reionization. 
Conversely, a bump model with $k_{\rm peak} = 7.3$
increases the number density of low-mass halos, producing 
a higher collapsed fraction and 
faster structure formation with earlier completion of reionization. 
Of particular interest is the ``turnover scale'' $k^{\rm turn} = 0.5$, which predicts 
almost the same number density of low-mass halos as the fiducial model. 
Consequently, this model shows negligible changes 
in the collapsed fraction and globally averaged quantities, as demonstrated in ref.~\cite{Naik:2025mba}.
The value of $k^{\rm turn}$, however, uniquely depends on the choice of 
$M_{\rm min}$ or $T_{\rm vir}$. 

\def\apj{ApJ}%
\def\mnras{MNRAS}%
\def\aap{A\&A}%
\def\apjl{ApJ}
\def\aj{AJ}
\def\physrep{PhR}
\def\apjs{ApJS}
\def\jcap{JCAP}
\def\pasa{PASA}
\def\pasj{PASJ}
\def\nat{Natur}
\def\apss{Ap\&SS}
\def\araa{ARA\&A}
\def\aaps{A\&AS}
\def\ssr{Space Sci. Rev.}
\def\pasp{PASP}
\def\na{New A}
\def\psj{PSJ}

% \begin{thebibliography}{99}
\bibliography{references.bib}

\providecommand{\href}[2]{#2}\begingroup\raggedright\begin{thebibliography}{10}

\bibitem{2024Natur.633..318C}
S.~{Carniani}, K.~{Hainline}, F.~{D'Eugenio}, D.J.~{Eisenstein}, P.~{Jakobsen}, J.~{Witstok} et~al., \emph{{Spectroscopic confirmation of two luminous galaxies at a redshift of 14}}, \href{https://doi.org/10.1038/s41586-024-07860-9}{\emph{\nat} {\bfseries 633} (2024) 318} [\href{https://arxiv.org/abs/2405.18485}{{\ttfamily 2405.18485}}].

\bibitem{Fan_2006}
X.~Fan, M.A.~Strauss, R.H.~Becker, R.L.~White, J.E.~Gunn, G.R.~Knapp et~al., \emph{Constraining the evolution of the ionizing background and the epoch of reionization with $z\sim 6$ quasars. ii. a sample of 19 quasars}, \href{https://doi.org/10.1086/504836}{\emph{The Astronomical Journal} {\bfseries 132} (2006) 117}.

\bibitem{Furlanetto:2006jb}
S.~Furlanetto, S.P.~Oh and F.~Briggs, \emph{{Cosmology at Low Frequencies: The 21 cm Transition and the High-Redshift Universe}}, \href{https://doi.org/10.1016/j.physrep.2006.08.002}{\emph{Phys. Rept.} {\bfseries 433} (2006) 181} [\href{https://arxiv.org/abs/astro-ph/0608032}{{\ttfamily astro-ph/0608032}}].

\bibitem{Pritchard:2011xb}
J.R.~Pritchard and A.~Loeb, \emph{{21-cm cosmology}}, \href{https://doi.org/10.1088/0034-4885/75/8/086901}{\emph{Rept. Prog. Phys.} {\bfseries 75} (2012) 086901} [\href{https://arxiv.org/abs/1109.6012}{{\ttfamily 1109.6012}}].

\bibitem{Chluba:2015bqa}
J.~Chluba, J.~Hamann and S.P.~Patil, \emph{{Features and New Physical Scales in Primordial Observables: Theory and Observation}}, \href{https://doi.org/10.1142/S0218271815300232}{\emph{Int. J. Mod. Phys. D} {\bfseries 24} (2015) 1530023} [\href{https://arxiv.org/abs/1505.01834}{{\ttfamily 1505.01834}}].

\bibitem{Zhao:2025ddy}
M.-L.~Zhao, Y.~Shao, S.~Wang and X.~Zhang, \emph{{Prospects for probing dark matter particles and primordial black holes with the Square Kilometre Array using the 21 cm power spectrum at cosmic dawn*}}, \href{https://doi.org/10.1088/1674-1137/ae1375}{\emph{Chin. Phys.} {\bfseries 50} (2026) 025101} [\href{https://arxiv.org/abs/2507.02651}{{\ttfamily 2507.02651}}].

\bibitem{Park:2025phj}
H.~Park, R.~Barkana, N.~Yoshida, S.~Sikder, R.~Mondal and A.~Fialkov, \emph{{The signature of subgalactic dark matter clumping in the global 21-cm signal of hydrogen}}, \href{https://doi.org/10.1038/s41550-025-02637-0}{\emph{Nature Astron.} {\bfseries 9} (2025) 1723} [\href{https://arxiv.org/abs/2509.11055}{{\ttfamily 2509.11055}}].

\bibitem{Cooray:2008eb}
A.~Cooray, C.~Li and A.~Melchiorri, \emph{{The trispectrum of 21-cm background anisotropies as a probe of primordial non-Gaussianity}}, \href{https://doi.org/10.1103/PhysRevD.77.103506}{\emph{Phys. Rev. D} {\bfseries 77} (2008) 103506} [\href{https://arxiv.org/abs/0801.3463}{{\ttfamily 0801.3463}}].

\bibitem{Pillepich:2006fj}
A.~Pillepich, C.~Porciani and S.~Matarrese, \emph{{The bispectrum of redshifted 21-cm fluctuations from the dark ages}}, \href{https://doi.org/10.1086/517963}{\emph{Astrophys. J.} {\bfseries 662} (2007) 1} [\href{https://arxiv.org/abs/astro-ph/0611126}{{\ttfamily astro-ph/0611126}}].

\bibitem{Munoz:2015eqa}
J.B.~Mu{\~n}oz, Y.~Ali-Ha{\"\i}moud and M.~Kamionkowski, \emph{{Primordial non-gaussianity from the bispectrum of 21-cm fluctuations in the dark ages}}, \href{https://doi.org/10.1103/PhysRevD.92.083508}{\emph{Phys. Rev. D} {\bfseries 92} (2015) 083508} [\href{https://arxiv.org/abs/1506.04152}{{\ttfamily 1506.04152}}].

\bibitem{Meerburg:2016zdz}
P.D.~Meerburg, M.~M{\"u}nchmeyer, J.B.~Mu{\~n}oz and X.~Chen, \emph{{Prospects for Cosmological Collider Physics}}, \href{https://doi.org/10.1088/1475-7516/2017/03/050}{\emph{JCAP} {\bfseries 03} (2017) 050} [\href{https://arxiv.org/abs/1610.06559}{{\ttfamily 1610.06559}}].

\bibitem{Floss:2022grj}
T.~Fl{\"o}ss, T.~de~Wild, P.D.~Meerburg and L.V.E.~Koopmans, \emph{{The Dark Ages' 21-cm trispectrum}}, \href{https://doi.org/10.1088/1475-7516/2022/06/020}{\emph{JCAP} {\bfseries 06} (2022) 020} [\href{https://arxiv.org/abs/2201.08843}{{\ttfamily 2201.08843}}].

\bibitem{Chung:1999ve}
D.J.H.~Chung, E.W.~Kolb, A.~Riotto and I.I.~Tkachev, \emph{{Probing Planckian physics: Resonant production of particles during inflation and features in the primordial power spectrum}}, \href{https://doi.org/10.1103/PhysRevD.62.043508}{\emph{Phys. Rev.} {\bfseries D62} (2000) 043508} [\href{https://arxiv.org/abs/hep-ph/9910437}{{\ttfamily hep-ph/9910437}}].

\bibitem{Barnaby:2009mc}
N.~Barnaby, Z.~Huang, L.~Kofman and D.~Pogosyan, \emph{{Cosmological Fluctuations from Infra-Red Cascading During Inflation}}, \href{https://doi.org/10.1103/PhysRevD.80.043501}{\emph{Phys. Rev.} {\bfseries D80} (2009) 043501} [\href{https://arxiv.org/abs/0902.0615}{{\ttfamily 0902.0615}}].

\bibitem{Barnaby:2009dd}
N.~Barnaby and Z.~Huang, \emph{{Particle Production During Inflation: Observational Constraints and Signatures}}, \href{https://doi.org/10.1103/PhysRevD.80.126018}{\emph{Phys. Rev. D} {\bfseries 80} (2009) 126018} [\href{https://arxiv.org/abs/0909.0751}{{\ttfamily 0909.0751}}].

\bibitem{Pearce:2017bdc}
L.~Pearce, M.~Peloso and L.~Sorbo, \emph{{Resonant particle production during inflation: a full analytical study}}, \href{https://doi.org/10.1088/1475-7516/2017/05/054}{\emph{JCAP} {\bfseries 1705} (2017) 054} [\href{https://arxiv.org/abs/1702.07661}{{\ttfamily 1702.07661}}].

\bibitem{Furuuchi:2015foh}
K.~Furuuchi, \emph{{Excursions through KK modes}}, \href{https://doi.org/10.1088/1475-7516/2016/07/008}{\emph{JCAP} {\bfseries 1607} (2016) 008} [\href{https://arxiv.org/abs/1512.04684}{{\ttfamily 1512.04684}}].

\bibitem{Furuuchi:2020klq}
K.~Furuuchi, S.S.~Naik and N.J.~Jobu, \emph{{Large Field Excursions from Dimensional (De)construction}}, \href{https://doi.org/10.1088/1475-7516/2020/06/054}{\emph{JCAP} {\bfseries 06} (2020) 054} [\href{https://arxiv.org/abs/2001.06518}{{\ttfamily 2001.06518}}].

\bibitem{Furuuchi:2020ery}
K.~Furuuchi, N.J.~Jobu and S.S.~Naik, \emph{{Extra-Natural Inflation (De)constructed}},  \href{https://arxiv.org/abs/2004.13755}{{\ttfamily 2004.13755}}.

\bibitem{Naik:2022mxn}
S.S.~Naik, K.~Furuuchi and P.~Chingangbam, \emph{{Particle production during inflation: a Bayesian analysis with CMB data from Planck 2018}}, \href{https://doi.org/10.1088/1475-7516/2022/07/016}{\emph{JCAP} {\bfseries 07} (2022) 016} [\href{https://arxiv.org/abs/2202.05862}{{\ttfamily 2202.05862}}].

\bibitem{Naik:2022wej}
S.S.~Naik, P.~Chingangbam and K.~Furuuchi, \emph{{Particle production during inflation: constraints expected from redshifted 21 cm observations from the epoch of reionization}}, \href{https://doi.org/10.1088/1475-7516/2023/04/058}{\emph{JCAP} {\bfseries 04} (2023) 058} [\href{https://arxiv.org/abs/2212.14064}{{\ttfamily 2212.14064}}].

\bibitem{Naik:2025mba}
S.S.~Naik, P.~Chingangbam, S.~Singh, A.~Mesinger and K.~Furuuchi, \emph{{Global 21 cm signal: a promising probe of primordial features}}, \href{https://doi.org/10.1088/1475-7516/2025/05/038}{\emph{JCAP} {\bfseries 05} (2025) 038} [\href{https://arxiv.org/abs/2501.02538}{{\ttfamily 2501.02538}}].

\bibitem{LOFAR2:2021A&A...652A..37E}
H.W.~{Edler}, F.~{de Gasperin} and D.~{Rafferty}, \emph{{Investigating ionospheric calibration for LOFAR 2.0 with simulated observations}}, \href{https://doi.org/10.1051/0004-6361/202140465}{\emph{\aap} {\bfseries 652} (2021) A37} [\href{https://arxiv.org/abs/2105.04636}{{\ttfamily 2105.04636}}].

\bibitem{MWA2:2018PASA...35...33W}
R.B.~{Wayth}, S.J.~{Tingay}, C.M.~{Trott}, D.~{Emrich}, M.~{Johnston-Hollitt}, B.~{McKinley} et~al., \emph{{The Phase II Murchison Widefield Array: Design overview}}, \href{https://doi.org/10.1017/pasa.2018.37}{\emph{\pasa} {\bfseries 35} (2018) e033} [\href{https://arxiv.org/abs/1809.06466}{{\ttfamily 1809.06466}}].

\bibitem{DeBoer:2016tnn}
D.R.~DeBoer et~al., \emph{{Hydrogen Epoch of Reionization Array (HERA)}}, \href{https://doi.org/10.1088/1538-3873/129/974/045001}{\emph{Publ. Astron. Soc. Pac.} {\bfseries 129} (2017) 045001} [\href{https://arxiv.org/abs/1606.07473}{{\ttfamily 1606.07473}}].

\bibitem{ska}
\url{http://www.skatelescope.org/}.

\bibitem{Bharadwaj:2004sx}
S.~Bharadwaj and S.K.~Pandey, \emph{{Probing non-Gaussian features in the HI distribution at the epoch of reionization}}, \href{https://doi.org/10.1111/j.1365-2966.2005.08836.x}{\emph{Mon. Not. Roy. Astron. Soc.} {\bfseries 358} (2005) 968} [\href{https://arxiv.org/abs/astro-ph/0410581}{{\ttfamily astro-ph/0410581}}].

\bibitem{Shimabukuro:2015iqa}
H.~Shimabukuro, S.~Yoshiura, K.~Takahashi, S.~Yokoyama and K.~Ichiki, \emph{{21 cm line bispectrum as a method to probe cosmic dawn and epoch of reionization}}, \href{https://doi.org/10.1093/mnras/stw482}{\emph{Mon. Not. Roy. Astron. Soc.} {\bfseries 458} (2016) 3003} [\href{https://arxiv.org/abs/1507.01335}{{\ttfamily 1507.01335}}].

\bibitem{Majumdar:2017tdm}
S.~Majumdar, J.R.~Pritchard, R.~Mondal, C.A.~Watkinson, S.~Bharadwaj and G.~Mellema, \emph{{Quantifying the non-Gaussianity in the EoR 21-cm signal through bispectrum}}, \href{https://doi.org/10.1093/mnras/sty535}{\emph{Mon. Not. Roy. Astron. Soc.} {\bfseries 476} (2018) 4007} [\href{https://arxiv.org/abs/1708.08458}{{\ttfamily 1708.08458}}].

\bibitem{Watkinson:2018efd}
C.A.~Watkinson, S.K.~Giri, H.E.~Ross, K.L.~Dixon, I.T.~Iliev, G.~Mellema et~al., \emph{{The 21-cm bispectrum as a probe of non-Gaussianities due to X-ray heating}}, \href{https://doi.org/10.1093/mnras/sty2740}{\emph{Mon. Not. Roy. Astron. Soc.} {\bfseries 482} (2019) 2653} [\href{https://arxiv.org/abs/1808.02372}{{\ttfamily 1808.02372}}].

\bibitem{Hutter:2019yta}
A.~Hutter, C.A.~Watkinson, J.~Seiler, P.~Dayal, M.~Sinha and D.J.~Croton, \emph{{The 21 cm bispectrum during reionization: a tracer of the ionization topology}}, \href{https://doi.org/10.1093/mnras/stz3139}{\emph{Mon. Not. Roy. Astron. Soc.} {\bfseries 492} (2020) 653} [\href{https://arxiv.org/abs/1907.04342}{{\ttfamily 1907.04342}}].

\bibitem{Noble:2024uzl}
L.~Noble, M.~Kamran, S.~Majumdar, C.S.~Murmu, R.~Ghara, G.~Mellema et~al., \emph{{Impact of the Epoch of Reionization sources on the 21-cm bispectrum}}, \href{https://doi.org/10.1088/1475-7516/2024/10/003}{\emph{JCAP} {\bfseries 10} (2024) 003} [\href{https://arxiv.org/abs/2406.03118}{{\ttfamily 2406.03118}}].

\bibitem{Tomita:1986}
H.~{Tomita}, \emph{{Curvature Invariants of Random Interface Generated by Gaussian Fields}}, \href{https://doi.org/10.1143/PTP.76.952}{\emph{Progress of Theoretical Physics} {\bfseries 76} (1986) 952}.

\bibitem{Mecke:1994}
K.R.~{Mecke}, T.~{Buchert} and H.~{Wagner}, \emph{{Robust morphological measures for large-scale structure in the Universe}}, \href{https://doi.org/10.48550/arXiv.astro-ph/9312028}{\emph{\aap} {\bfseries 288} (1994) 697} [\href{https://arxiv.org/abs/astro-ph/9312028}{{\ttfamily astro-ph/9312028}}].

\bibitem{Schmalzing:1997}
J.~{Schmalzing} and T.~{Buchert}, \emph{{Beyond Genus Statistics: A Unifying Approach to the Morphology of Cosmic Structure}}, \href{https://doi.org/10.1086/310680}{\emph{\apjl} {\bfseries 482} (1997) L1} [\href{https://arxiv.org/abs/astro-ph/9702130}{{\ttfamily astro-ph/9702130}}].

\bibitem{Matsubara:2003}
T.~Matsubara, \emph{Statistics of smoothed cosmic fields in perturbation theory. i. formulation and useful formulae in second-order perturbation theory}, \href{https://doi.org/10.1086/345521}{\emph{The Astrophysical Journal} {\bfseries 584} (2003) 1}.

\bibitem{Gleser_2006}
L.~Gleser, A.~Nusser, B.~Ciardi and V.~Desjacques, \emph{The morphology of cosmological reionization by means of minkowski functionals: Morphology of reionization}, \href{https://doi.org/10.1111/j.1365-2966.2006.10556.x}{\emph{Monthly Notices of the Royal Astronomical Society} {\bfseries 370} (2006) 1329–1338}.

\bibitem{yoshiura2016}
S.~Yoshiura, H.~Shimabukuro, K.~Takahashi and T.~Matsubara, \emph{Studying topological structure of 21-cm line fluctuations with 3d minkowski functionals before reionization}, \href{https://doi.org/10.1093/mnras/stw2701}{\emph{Monthly Notices of the Royal Astronomical Society} {\bfseries 465} (2016) 394} [\href{https://arxiv.org/abs/https://academic.oup.com/mnras/article-pdf/465/1/394/8593179/stw2701.pdf}{{\ttfamily https://academic.oup.com/mnras/article-pdf/465/1/394/8593179/stw2701.pdf}}].

\bibitem{Diao_2024}
K.~Diao, Z.~Chen, X.~Chen and Y.~Mao, \emph{Reionization parameter inference from 3d minkowski functionals of the 21 cm signals}, \href{https://doi.org/10.3847/1538-4357/ad6c40}{\emph{The Astrophysical Journal} {\bfseries 974} (2024) 141}.

\bibitem{Bag:2018}
S.~{Bag}, R.~{Mondal}, P.~{Sarkar}, S.~{Bharadwaj} and V.~{Sahni}, \emph{{The shape and size distribution of H II regions near the percolation transition}}, \href{https://doi.org/10.1093/mnras/sty714}{\emph{\mnras} {\bfseries 477} (2018) 1984} [\href{https://arxiv.org/abs/1801.01116}{{\ttfamily 1801.01116}}].

\bibitem{Bag:2019}
S.~{Bag}, R.~{Mondal}, P.~{Sarkar}, S.~{Bharadwaj}, T.R.~{Choudhury} and V.~{Sahni}, \emph{{Studying the morphology of H I isodensity surfaces during reionization using Shapefinders and percolation analysis}}, \href{https://doi.org/10.1093/mnras/stz532}{\emph{\mnras} {\bfseries 485} (2019) 2235} [\href{https://arxiv.org/abs/1809.05520}{{\ttfamily 1809.05520}}].

\bibitem{Ghara:2023efi}
R.~Ghara, S.~Bag, S.~Zaroubi and S.~Majumdar, \emph{{The morphology of the redshifted 21-cm signal from the Cosmic Dawn}}, \href{https://doi.org/10.1093/mnras/stae895}{\emph{Mon. Not. Roy. Astron. Soc.} {\bfseries 530} (2024) 191} [\href{https://arxiv.org/abs/2308.00548}{{\ttfamily 2308.00548}}].

\bibitem{Lee_2008}
K.-G.~Lee, R.~Cen, J.R.~Gott~III and H.~Trac, \emph{The topology of cosmological reionization}, \href{https://doi.org/10.1086/525520}{\emph{The Astrophysical Journal} {\bfseries 675} (2008) 8}.

\bibitem{Ganesan:2017}
V.~{Ganesan} and P.~{Chingangbam}, \emph{{Tensor Minkowski Functionals: first application to the CMB}}, \href{https://doi.org/10.1088/1475-7516/2017/06/023}{\emph{\jcap} {\bfseries 2017} (2017) 023} [\href{https://arxiv.org/abs/1608.07452}{{\ttfamily 1608.07452}}].

\bibitem{Chingangbam:2017}
P.~{Chingangbam}, K.P.~{Yogendran}, P.K.~{Joby}, V.~{Ganesan}, S.~{Appleby} and C.~{Park}, \emph{{Tensor Minkowski Functionals for random fields on the sphere}}, \href{https://doi.org/10.1088/1475-7516/2017/12/023}{\emph{\jcap} {\bfseries 2017} (2017) 023} [\href{https://arxiv.org/abs/1707.04386}{{\ttfamily 1707.04386}}].

\bibitem{Appleby:2018a}
S.~{Appleby}, P.~{Chingangbam}, C.~{Park}, S.E.~{Hong}, J.~{Kim} and V.~{Ganesan}, \emph{{Minkowski Tensors in Two Dimensions: Probing the Morphology and Isotropy of the Matter and Galaxy Density Fields}}, \href{https://doi.org/10.3847/1538-4357/aabb53}{\emph{\apj} {\bfseries 858} (2018) 87} [\href{https://arxiv.org/abs/1712.07466}{{\ttfamily 1712.07466}}].

\bibitem{Appleby:2018b}
S.~{Appleby}, P.~{Chingangbam}, C.~{Park}, K.P.~{Yogendran} and P.K.~{Joby}, \emph{{Minkowski Tensors in Three Dimensions: Probing the Anisotropy Generated by Redshift Space Distortion}}, \href{https://doi.org/10.3847/1538-4357/aacf8c}{\emph{\apj} {\bfseries 863} (2018) 200} [\href{https://arxiv.org/abs/1805.08752}{{\ttfamily 1805.08752}}].

\bibitem{Kapahtia:2017qrg}
A.~Kapahtia, P.~Chingangbam, S.~Appleby and C.~Park, \emph{{A novel probe of ionized bubble shape and size statistics of the epoch of reionization using the contour Minkowski Tensor}}, \href{https://doi.org/10.1088/1475-7516/2018/10/011}{\emph{JCAP} {\bfseries 10} (2018) 011} [\href{https://arxiv.org/abs/1712.09195}{{\ttfamily 1712.09195}}].

\bibitem{Kapahtia_2019}
A.~Kapahtia, P.~Chingangbam and S.~Appleby, \emph{Morphology of 21cm brightness temperature during the epoch of reionization using contour minkowski tensor}, \href{https://doi.org/10.1088/1475-7516/2019/09/053}{\emph{Journal of Cosmology and Astroparticle Physics} {\bfseries 2019} (2019) 053}.

\bibitem{Kapahtia:2021eok}
A.~Kapahtia, P.~Chingangbam, R.~Ghara, S.~Appleby and T.R.~Choudhury, \emph{{Prospects of constraining reionization model parameters using Minkowski tensors and Betti numbers}}, \href{https://doi.org/10.1088/1475-7516/2021/05/026}{\emph{JCAP} {\bfseries 05} (2021) 026} [\href{https://arxiv.org/abs/2101.03962}{{\ttfamily 2101.03962}}].

\bibitem{Park:2013}
C.~{Park}, P.~{Pranav}, P.~{Chingangbam}, R.~{Van De Weygaert}, B.~{Jones}, G.~{Vegter} et~al., \emph{{BETTI Numbers of Gaussian Fields}}, \href{https://doi.org/10.5303/JKAS.2013.46.3.125}{\emph{Journal of Korean Astronomical Society} {\bfseries 46} (2013) 125} [\href{https://arxiv.org/abs/1307.2384}{{\ttfamily 1307.2384}}].

\bibitem{Giri_2021}
S.K.~Giri and G.~Mellema, \emph{Measuring the topology of reionization with betti numbers}, \href{https://doi.org/10.1093/mnras/stab1320}{\emph{Monthly Notices of the Royal Astronomical Society} {\bfseries 505} (2021) 1863–1877}.

\bibitem{2019MNRAS.486.1523E}
W.~{Elbers} and R.~{van de Weygaert}, \emph{{Persistent topology of the reionization bubble network - I. Formalism and phenomenology}}, \href{https://doi.org/10.1093/mnras/stz908}{\emph{\mnras} {\bfseries 486} (2019) 1523} [\href{https://arxiv.org/abs/1812.00462}{{\ttfamily 1812.00462}}].

\bibitem{Elbers_2023}
W.~Elbers and R.~van de Weygaert, \emph{Persistent topology of the reionization bubble network – ii. evolution and classification}, \href{https://doi.org/10.1093/mnras/stad120}{\emph{Monthly Notices of the Royal Astronomical Society} {\bfseries 520} (2023) 2709–2726}.

\bibitem{Mesinger:2011}
A.~{Mesinger}, S.~{Furlanetto} and R.~{Cen}, \emph{{21CMFAST: a fast, seminumerical simulation of the high-redshift 21-cm signal}}, \href{https://doi.org/10.1111/j.1365-2966.2010.17731.x}{\emph{Mon. Not. Roy. Astron. Soc.} {\bfseries 411} (2011) 955} [\href{https://arxiv.org/abs/1003.3878}{{\ttfamily 1003.3878}}].

\bibitem{Murray:2020trn}
S.G.~Murray, B.~Greig, A.~Mesinger, J.B.~Mu\~noz, Y.~Qin, J.~Park et~al., \emph{{21cmFAST v3: A Python-integrated C code for generating 3D realizations of the cosmic 21cm signal}}, \href{https://doi.org/10.21105/joss.02582}{\emph{J. Open Source Softw.} {\bfseries 5} (2020) 2582} [\href{https://arxiv.org/abs/2010.15121}{{\ttfamily 2010.15121}}].

\bibitem{Furlanetto_2006}
S.R.~Furlanetto, S.~Peng~Oh and F.H.~Briggs, \emph{Cosmology at low frequencies: The 21cm transition and the high-redshift universe}, \href{https://doi.org/10.1016/j.physrep.2006.08.002}{\emph{Physics Reports} {\bfseries 433} (2006) 181–301}.

\bibitem{Wouthuysen1958PIRE...46..240F}
G.B.~{Field}, \emph{{Excitation of the Hydrogen 21-CM Line}}, \href{https://doi.org/10.1109/JRPROC.1958.286741}{\emph{Proceedings of the IRE} {\bfseries 46} (1958) 240}.

\bibitem{Field1952AJ.....57R..31W}
S.A.~{Wouthuysen}, \emph{{On the excitation mechanism of the 21-cm (radio-frequency) interstellar hydrogen emission line.}}, \href{https://doi.org/10.1086/106661}{\emph{\aj} {\bfseries 57} (1952) 31}.

\bibitem{Hikage_2006}
C.~Hikage, E.~Komatsu and T.~Matsubara, \emph{Primordial non‐gaussianity and analytical formula for minkowski functionals of the cosmic microwave background and large‐scale structure}, \href{https://doi.org/10.1086/508653}{\emph{The Astrophysical Journal} {\bfseries 653} (2006) 11–26}.

\bibitem{Chen_2019}
Z.~Chen, Y.~Xu, Y.~Wang and X.~Chen, \emph{Stages of reionization as revealed by the minkowski functionals}, \href{https://doi.org/10.3847/1538-4357/ab43e6}{\emph{The Astrophysical Journal} {\bfseries 885} (2019) 23}.

\bibitem{Mesinger:2007pd}
A.~Mesinger and S.~Furlanetto, \emph{{Efficient Simulations of Early Structure Formation and Reionization}}, \href{https://doi.org/10.1086/521806}{\emph{Astrophys. J.} {\bfseries 669} (2007) 663} [\href{https://arxiv.org/abs/0704.0946}{{\ttfamily 0704.0946}}].

\bibitem{Bashir:2025}
M.~{Bashir}, N.~{S}, P.~{Chingangbam}, F.~{Rahman}, P.~{Goyal}, S.~{Appleby} et~al., \emph{{Local patch analysis of ACT DR6 convergence map using morphological statistics}}, \href{https://doi.org/10.48550/arXiv.2503.17849}{\emph{arXiv e-prints} (2025) arXiv:2503.17849} [\href{https://arxiv.org/abs/2503.17849}{{\ttfamily 2503.17849}}].

\bibitem{Press_Schechter1974ApJ...187..425P}
W.H.~{Press} and P.~{Schechter}, \emph{{Formation of Galaxies and Clusters of Galaxies by Self-Similar Gravitational Condensation}}, \href{https://doi.org/10.1086/152650}{\emph{\apj} {\bfseries 187} (1974) 425}.

\bibitem{Sheth:2001dp}
R.K.~Sheth and G.~Tormen, \emph{{An Excursion Set Model of Hierarchical Clustering : Ellipsoidal Collapse and the Moving Barrier}}, \href{https://doi.org/10.1046/j.1365-8711.2002.04950.x}{\emph{Mon. Not. Roy. Astron. Soc.} {\bfseries 329} (2002) 61} [\href{https://arxiv.org/abs/astro-ph/0105113}{{\ttfamily astro-ph/0105113}}].

\end{thebibliography}\endgroup
\bibliographystyle{JHEP}
% \end{thebibliography}

\end{document}